\documentclass[review]{elsarticle}

\usepackage{lineno,hyperref}
\usepackage{array}
\usepackage[normalem]{ulem}   
\usepackage{caption}
\usepackage{longtable}
\usepackage{multirow}
\usepackage[caption=false]{subfig}
\usepackage[table]{xcolor}
\usepackage{graphicx}

\biboptions{numbers,sort&compress}

\modulolinenumbers[5]

\journal{Computational Materials Science}

\begin{document}

\begin{frontmatter}

\title{Breathing bands due to molecular order in CH$_3$NH$_3$PbI$_3$}

\author{Ma\l{}gorzata~Wierzbowska}
\address{Institute of High Pressure Physics, Polish Academy of Sciences, Soko\l{}owska 29/37, 01-142, Warsaw, Poland}

\author{Juan Jos\'e Mel\'endez\fnref{mycorrespondingauthor}}
\address{Department of Physics, University of Extremadura, Avenida de Elvas, s/n, 06006, Badajoz, Spain}
\address{Institute for Advanced Scientific Computing of Extremadura
(ICCAEX), Avenida de Elvas, s/n, 06006, Badajoz, Spain}
\fntext[mycorrespondingauthor]{Corresponding author}

\author{Daniele Varsano}
\address{Center S3, CNR Institute of Nanoscience, Via Campi 213A, 41125 Modena, Italy}

\begin{abstract}
CH$_3$NH$_3$PbI$_3$ perovskite is nowadays amongst the most promising photovoltaic materials for energy conversion. We have studied by {\it ab-initio} calculations, using several levels of approximation \--namely density functional theory including spin-orbit coupling and quasi-particle corrections by means of the GW method, as well as pseudopotential self-interaction corrections\--, the role of the methylammonium orientation on the electronic structure of this perovskite. We have considered many molecular arrangements within $2\times2\times2$ supercells, showing that the relative orientation of the organic molecules is responsible for a huge band gap variation up to 2~eV. The band gap sizes are related to distortions of the PbI$_3$ cage, which are in turn due to electrostatic interactions between this inorganic frame and the molecules. The strong dependence of the band gap on the mutual molecular orientation is confirmed at all levels of approximations. Our results suggest then that the coupling between the molecular motion and the interactions of the molecules with the inorganic cage could help to explain the widening of the absorption spectrum of CH$_3$NH$_3$PbI$_3$ perovskite, consistent with the observed white spectrum.
\end{abstract}


\end{frontmatter}


\section{Introduction}

Organic-inorganic hybrid semiconductors arose in the 90's as unconventional 
semiconductors with promising optical and electronic properties \cite{Papavassiliou-97}. 
In a very general sense, they consist of an inorganic frame, which is the active 
(i.e., semiconducting) part, and of an organic part which is assumed to be passive and 
to hinder charge transport. The organic and inorganic parts affect the optical properties 
of these systems, since they may give rise respectively to Frenkel-type or 
to Wannier-type excitonic bands, and therefore to many features in the absorption 
spectra \cite{Papavassiliou-99}. The organic-inorganic hybrid semiconductors are nowadays 
used as elements in devices for optics or optoelectronics, including LEDs and, 
especially, solar cells \cite{Kim-15, Wright-12}.

\begin{figure}
\centering
\includegraphics[scale=0.45]{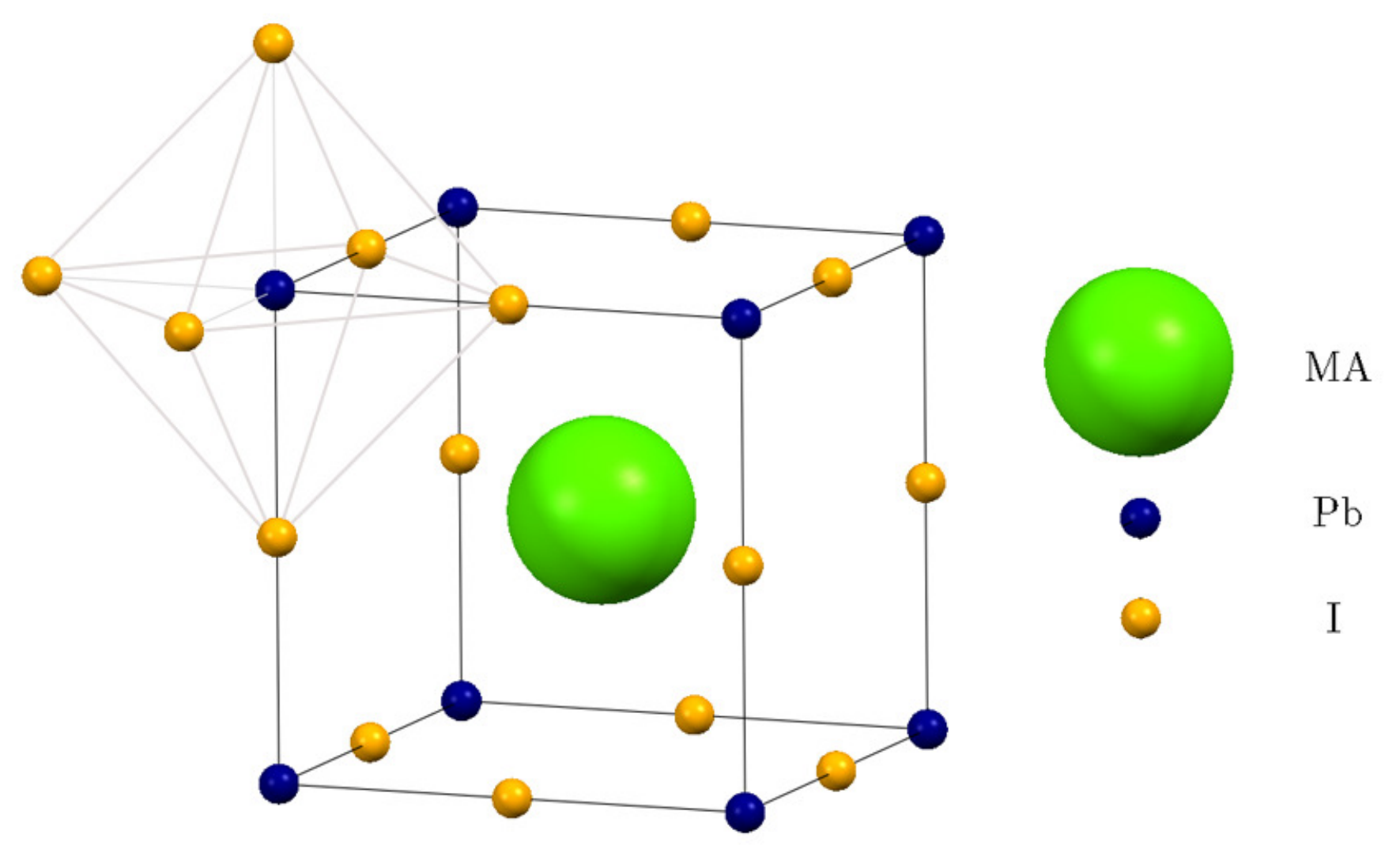}
\caption{Schematic representation of the perovskite structure of cubic MAPbI$_3$. 
The green circle represents the location of the MA molecule regardless its orientation.}
\label{F1}
\end{figure}

Among the myriad of hybrid semiconductors, methylammonium lead halide (CH$_3$NH$_3$PbI$_3$) 
has stood out during the last few years because of its excellent electronic and optical 
properties \cite{Kitazawa-02, Green-14}. 
Other materials in this family include in their composition formamidinium CH(NH$_2$)$_2^+$ 
instead of methylammonium, Sn in place of Pb, or Br, Cl or BF$_4$ 
replacing I \cite{Bisquert-16,Jeon-15}. CH$_3$NH$_3$PbI$_3$ crystallizes in the orthorhombic 
system at temperatures below around 162~K, becoming tetragonal at room temperature and 
cubic above 327~K, that is, for most temperatures of technological interest \cite{Baikie-13}. 
In this phase, it takes a perovskite-like structure 
with the methylammonium (MA) CH$_3$NH$_3$ molecule located in the centers of the cubes and 
the Pb and I atoms at the corners and centers of the edges, respectively, as schematized in Figure~\ref{F1}.
The MA molecules do not have a fixed orientation, as can be understood by 
symmetry considerations. Indeed, the methylammonium molecules have $C_{3v}$ symmetry, 
and compatibility with the $O_h$ symmetry of the perovskite structure requires them to be 
distributed at eight $(x,x,x)$ positions with random orientations of their C-N axes, 
as has been observed experimentally \cite{Leguy-16b,Kojima-09,Baikie-13,Wasylishen-85}.
In particular, the structural coherence length for the disordered structure has been estimated 
to be about 14 \AA, which matches the length of two PbI$_3$ cells \cite{Choi-14}. 
The MA cations are not only randomly distributed, but also able to perform 
rapid rotations, partially hindered as the temperature decreases \cite{Wasylishen-85}. 
The disordered and rapidly rotating nature of the MA molecules within MAPbI$_3$ 
is consistent with its apparent non-polar character \cite{Yamada-15,Mahale-16}, although 
this is still a controversial issue \cite{Frost-14}.

Kojima \textit{et al.} reported for the first time the effectiveness of 
MAPbI$_3$ - TiO$_2$ for visible-light conversion, with photovoltages 
close to 1.0~eV \cite{Kojima-09}. The energy conversion efficiency is more than 
the 15\% so far, larger than those of organic photovoltaics or dye-sensitized 
solar cells \cite{Green-14, Liu-13, Burschka-13}; some works even claim it to 
be well above the 20\% \cite{Gao-14,Bisquert-16}. Such a high photoconversion efficiency has been ascribed to its large absorption coefficient and to long range charge-carrier diffusion lengths~\cite{Xing-13,Giorgi-13}.
This justifies the hope arose by these materials for photovoltaic applications, although some issues, such as surface degradation under working conditions, are still under development \cite{Sowa}.

The optical absorption spectrum of MAPbI$_3$ reported so far ranges 
between about 350 to 800 nm (1.55--3.55~eV  approximately) \cite{Anand-16,Stoumpos-13}.
At room temperature, it consists of a broad band and does not exhibit peaks 
reflecting band structure details ~\cite{aharon-16,Even-13}.
Calculations within the density functional theory (DFT) and the 
many-body perturbation theory (MBPT) are consistent with electronic gaps ranging 
between 1.2 and 1.7~eV \cite{Baikie-13,Umebayashi-03,mosconi-13,motta-15,Umari-14,Amat-14,Zhu-14}, 
which agree reasonably well with photoemission experiments~\cite{Schulz-14} 
and with the onset of the optical absorption spectrum.
DFT calculations have also pointed out that the electronic gap width depends on 
the particular orientation of the MA molecules; in particular, larger gaps seem 
to correspond to systems with lower symmetry in most cases. 
This feature, which is common to other hybrid 
perovskites \cite{Borriello-08}, has been confirmed by experiments \cite{Baikie-13}.
In this sense, since the MA molecules are randomly oriented within 
the inorganic cage and they are free to rotate, one may consider 
that MAPbI$_3$ has a variable gap, which becomes wider or narrower 
depending on the particular arrangement of the MA molecules. 
This is relevant for optical properties, since the correspondence between 
the symmetry and the gap width theoretically allows some degree of 
``gap engineering'', which could ultimately yield tailored materials 
with optimized optical performance. 

DFT calculations showed that changes in the symmetry affect the direct or 
indirect nature of the electronic gap as well. In particular a direct gap located 
at $R$ is found when the MA molecules are all oriented along the $\langle 100 \rangle$ or 
$\langle 111 \rangle$ directions, whereas it becomes indirect when they align along 
$\langle 110 \rangle$ \cite{Amat-14,motta-15}; this feature 
is independent of the inclusion or not of spin-orbit coupling terms~\cite{motta-15}. 
The direct-indirect gap transition is relevant, since an indirect
band gap is responsible for slower electron-hole recombination \cite{Hutter-nmat}.
In relation to this, \textit{ab initio} 
calculations \cite{Mashiyama-98,motta-15,Amat-14}, as well as optical absorption 
measurements \cite{Leguy-16}, have pointed out that the inorganic cage distorts for some 
orientations of the MA molecule, which yields the direct-indirect gap transition. This suggests that the interactions between the cage 
and either the CH$_3$ or NH$_3$ groups of MA are in the end responsible for 
the nature and width of the band gap, and hence for the optical behavior, of MAPbI$_3$. 
 In this respect, the correlation between the optical band gap of metal-halide perovskites and metal–halide–metal bond angle was recently revealed by Filip {\it et al.} \cite{Filip-14}.

Most of the computational studies on the electronic structure of MAPbI$_3$ have focused 
on a single orientation of the MA molecule in a framework of periodic boundary conditions, 
which yields a solid with all the MA parallel to each other and oriented along 
a given direction. This picture, although computationally simple, 
is incompatible with the symmetry requirements as it does not take into 
account different arrangements of neighbor molecules. In this work 
we aim to study the interplay between the orientation of the molecules 
and the distortions of the cage, 
as well as its impact on the electronic structure of cubic MAPbI$_3$, 
by considering different patterns of molecular orientations in a supercell approach. 
The electronic structure calculations on these 
models were carried out at different levels of approximation, namely DFT with and 
without spin-orbit coupling corrections, pseudopotential self-interaction corrections and 
MBPT within the GW approximation. We show that DFT gaps are 
in good agreement with experimental data only fortuitously; recovering gaps comparable 
with experiments requires including many-body effects, as already pointed out in 
Refs~\cite{Umari-14,Ahmed-15,Zhu-14}. Besides, we show 
that the gap size variations may be correlated to interactions between 
the organic and the inorganic moieties of the system, which give rise to the aforementioned 
structural distortions. Finally, we show that there might exist some molecular arrangements 
capable to drastically increase or reduce the band gaps, 
thus modifying the optical performance of the system.

The paper is organized as follows: in Sec. \ref{tech}, we provide the computational details of our calculations (\ref{tech1}) and describe the adopted geometries for the model systems (\ref{tech2}). In Sec. \ref{results}, we present and analyze the results obtained for the structure and electronic properties of the different geometries considered. In Sec. \ref{dft1}, we describe structural properties and formation energies of the studied systems. The electronic properties at DFT level are reported in Sec. \ref{dft2}, and the band gap variations are discussed in Sec. \ref{dft3}. 
In Sec. \ref{SOC}, we discuss the inclusion of spin-orbit coupling, self interaction correction and many-body effects on the electronic structure of some selected configurations. Finally in Sec.~\ref{conclusion}, we present our conclusions and briefly discuss some implications that our results could have on the optical properties of the system.

\begin{figure}[!ht]
	\centering
	\subfloat[Case 1]{\label{F2a}\includegraphics[width=0.5\textwidth]{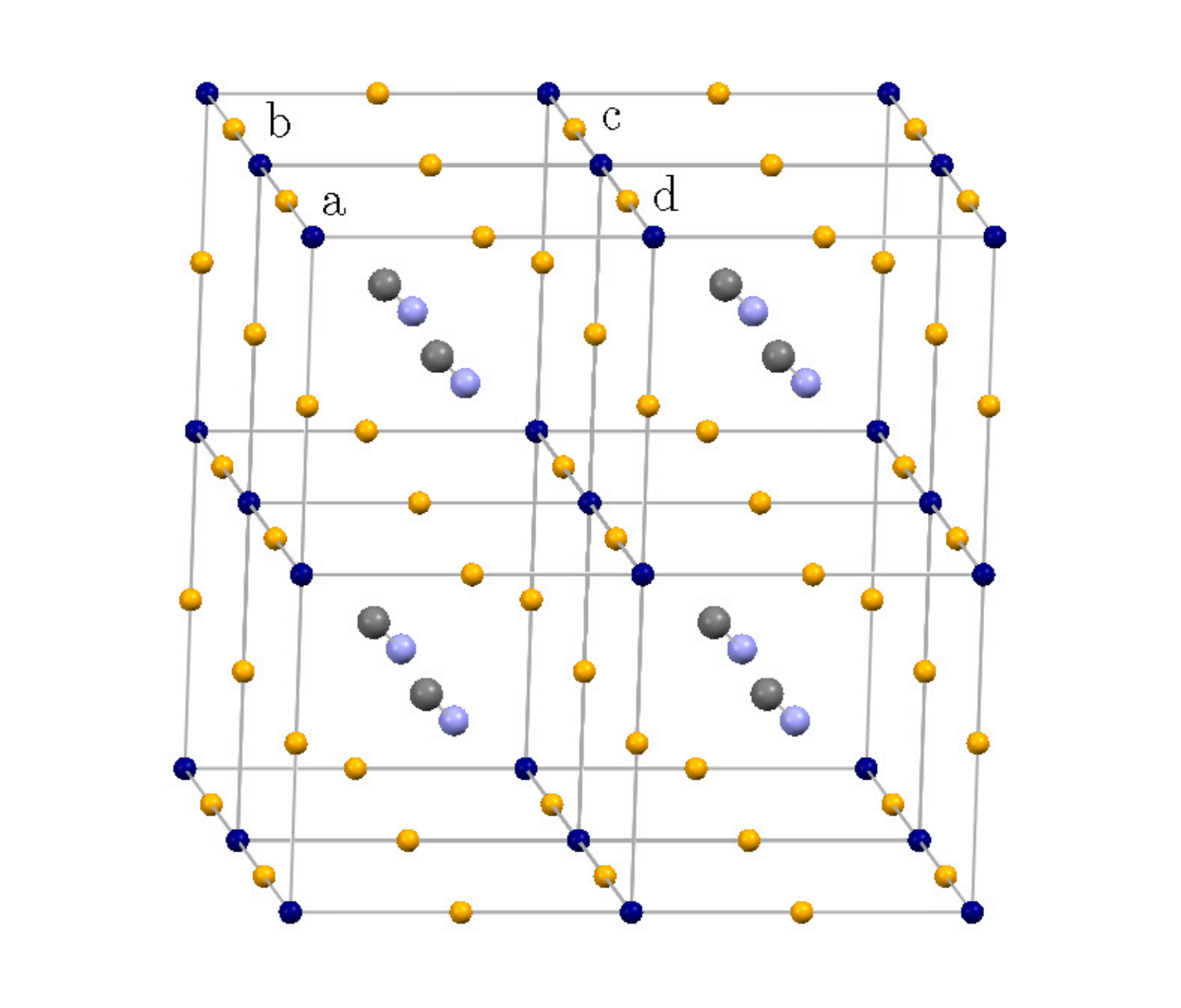}} 
	\subfloat[Case 15]{\label{F2b}\includegraphics[width=0.47\textwidth]{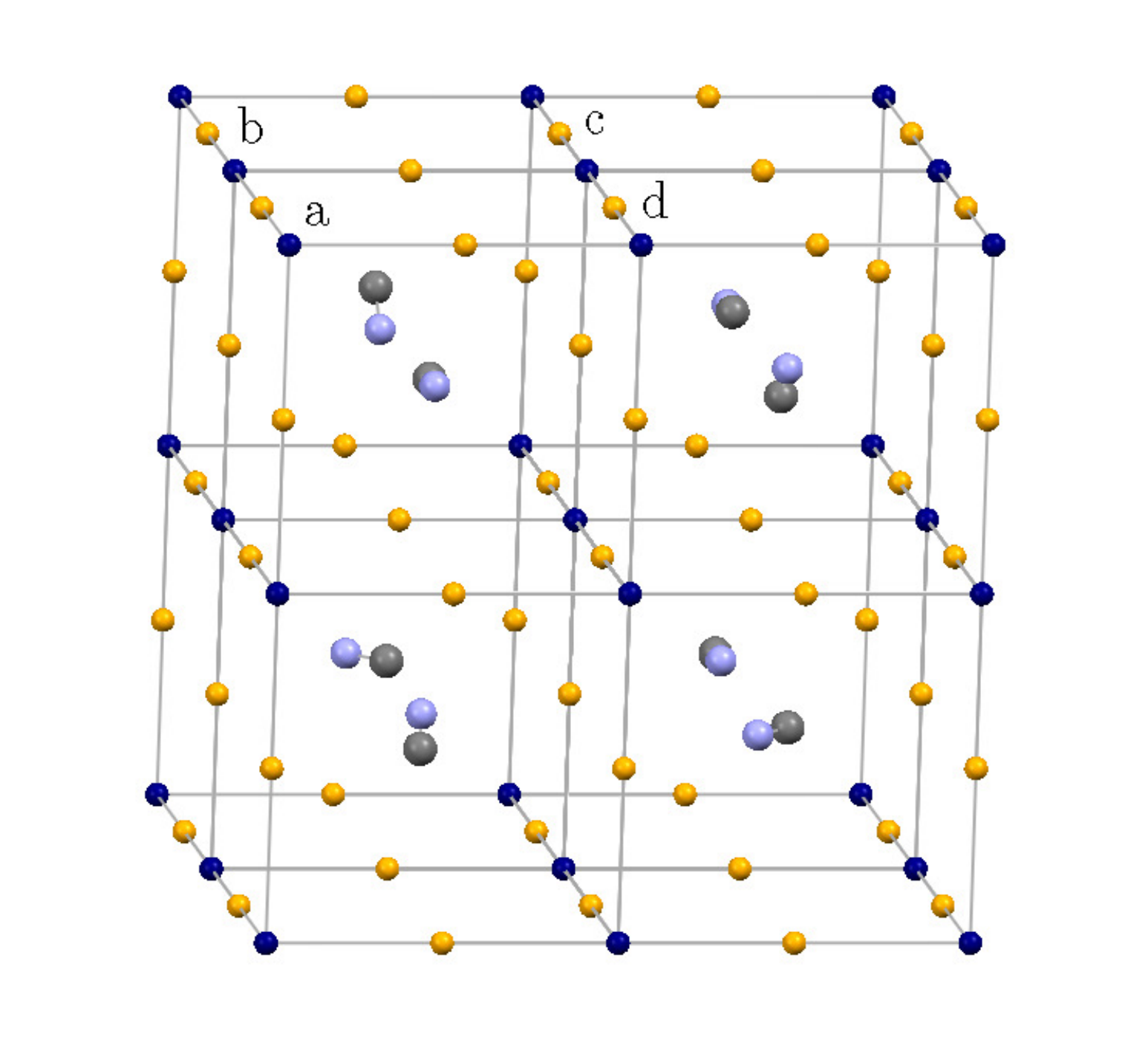}} 	
	\caption{Geometries of two selected cases taken from Table~\ref{configurations}.}
	\label{F2}
\end{figure}

\begin{table}[ht]
\begin{center}
\caption{The chosen molecular configurations for the 2$\times$2$\times$2 supercell. The a, b, c and d entries for each table denote four MAPbI$_3$ unit cells taken in clockwise sense in the upper and bottom plane of the 2$\times$2$\times$2 supercell, as outlined in Figure~\ref{F2}.}
\vspace{1mm}
\begin{tabular}{c | cccc | cccc} 
\hline  
\multirow{2}{*}{Case} & \multicolumn{4}{c|}{Upper plane} & \multicolumn{4}{c}{Bottom plane} \\[0.1cm] \cline{2-9}
& a & b & c & d & a & b & c & d \\ 
 \hline

 1 & $[111]$ & $[111]$ & $[111]$ & $[111]$ & $[111]$ & $[111]$ & $[111]$ & $[111]$ \\
 2 & $[110]$ & $[110]$ & $[110]$ & $[110]$ & $[110]$ & $[110]$ & $[110]$ & $[110]$ \\
 3 & $[100]$ & $[100]$ & $[100]$ & $[100]$ & $[100]$ & $[100]$ & $[100]$ & $[100]$ \\
 4 & $[111]$ & $[\bar1\bar1\bar1]$ & $[111]$ & $[\bar1\bar1\bar1]$ & $[111]$ & $[\bar1\bar1\bar1]$ & $[111]$ & $[\bar1\bar1\bar1]$ \\
 5 & $[111]$ & $[\bar 111]$ & $[\bar 111]$ & $[111]$ & $[111]$ & $[\bar 111]$ & $[\bar 111]$ & $[111]$ \\
 6 & $[011]$ & $[0\bar1\bar1]$ & $[0\bar1\bar1]$ & $[011]$ & $[011]$ & $[0\bar1\bar1]$ & $[0\bar1\bar1]$ & $[011]$ \\
 7 & $[111]$ & $[\bar111]$ & $[\bar111]$ & $[111]$ & $[111]$ & $[\bar 111]$ & $[\bar111]$ & $[111]$ \\
 8 & $[1\bar11]$ & $[\bar1\bar11]$ & $[\bar111]$ & $[111]$ & $[1\bar11]$ & $[\bar1\bar11]$ & $[\bar111]$ & $[111]$ \\
 9 & $[001]$ & $[00\bar1]$ & $[001]$ & $[00\bar1]$ & $[001]$ & $[00\bar1]$ & $[001]$ & $[00\bar1]$ \\
 10 & $[001]$ & $[00\bar1]$ & $[001]$ & $[00\bar1]$ & $[00\bar1]$ & $[001]$ & $[00\bar1]$ & $[001]$ \\
 11 & $[111]$ & $[\bar111]$ & $[\bar1\bar11]$ & $[1\bar11]$ & $[\bar111]$ & $[111]$ & $[\bar111]$ & $[111]$ \\
 12 & $[010]$ & $[010]$ & $[0\bar10]$ & $[110]$ & $[0\bar10]$ & $[\bar100]$ & $[100]$ & $[\bar100]$ \\
 13 & $[001]$ & $[0\bar10]$ & $[00\bar1]$ & $[010]$ & $[0\bar10]$ & $[00\bar1]$ & $[010]$ & $[001]$ \\ 
 14 & $[00\bar1]$ & $[001]$ & $[00\bar1]$ & $[100]$ & $[100]$ & $[\bar100]$ & $[\bar100]$ & $[001]$ \\
 15 & $[\bar1\bar10]$ & $[010]$ & $[0\bar10]$ & $[0\bar1\bar1]$ & $[0\bar10]$ & $[1\bar10]$ & $[01\bar1]$ & $[\bar100]$ \\
 16 & $[\bar1\bar10]$ & $[011]$ & $[110]$ & $[0\bar11]$ & $[0\bar1\bar1]$ & $[1\bar10]$ & $[01\bar1]$ & $[\bar110]$ \\
 17 & $[10\bar1]$ & $[\bar110]$ & $[\bar10\bar1]$ & $[1\bar10]$ & $[110]$ & $[\bar101]$ & $[\bar1\bar10]$ & $[101]$ \\
 18 & $[\bar101]$ & $[0\bar1\bar1]$ & $[101]$ & $[01\bar1]$ & $[0\bar11]$ & $[10\bar1]$ & $[011]$ & $[\bar10\bar1]$ \\
 19 & $[\bar1\bar1\bar1]$ & $[11\bar1]$ & $[11\bar1]$ & $[\bar1\bar1\bar1]$ & $[\bar111]$ & $[\bar111]$ & $[1\bar11]$ & $[1\bar11]$ \\
 20 & $[\bar11\bar1]$ & $[\bar111]$ & $[\bar1\bar1\bar1]$ & $[11\bar1]$ & $[\bar1\bar11]$ & $[1\bar11]$ & $[1\bar1\bar1]$ & $[111]$ \\
 21 & $[\bar11\bar1]$ & $[1\bar1\bar1]$ & $[1\bar1\bar1]$ & $[\bar1\bar11]$ & $[\bar11\bar1]$ & $[111]$ & $[111]$ & $[\bar1\bar11]$ \\
 22 & $[001]$ & $[0\bar10]$ & $[\bar1\bar1\bar1]$ & $[1\bar10]$ & $[\bar1\bar11]$ & $[\bar101]$ & $[010]$ & $[0\bar10]$ \\
 23 & $[\bar1\bar10]$ & $[010]$ & $[110]$ & $[\bar1\bar1\bar1]$ & $[0\bar1\bar1]$ & $[\bar111]$ & $[01\bar1]$ & $[\bar100]$ \\
 24 & $[\bar11\bar1]$ & $[1\bar1\bar1]$ & $[00\bar1]$ & $[\bar1\bar11]$ & $[100]$ & $[101]$ & $[111]$ & $[\bar1\bar11]$ \\
   \hline
\end{tabular}
\label{configurations}
\end{center}
\end{table}

\section{Theoretical methods}\label{tech}
\subsection{Computational details}\label{tech1}

DFT calculations were performed using the plane-wave QUANTUM ESPRESSO (QE) package \cite{giannozzi-09} within the generalized gradient approximation (GGA), with the Perdew-Burke-Ernzerhof parametrization for 
the exchange-correlation functional \cite{Perdew-96}.
Norm-conserving pseudopotentials were used to model the electron-ion interaction and the kinetic energy cutoff for the wave functions was set to 60~Ry.
Due to the presence of Pb atoms, we also took into account 
the spin-orbit coupling (SOC) effect and, therefore, two sets of 
pseudopotentials were used: scalar relativistic (DFT without SOC) and 
full relativistic (SOC-DFT). 
In all cases, the first Brillouin zone was sampled with a 
4$\times$4$\times$4 Monkhorst-Pack grid \cite{Monkhorst-76}.

To go beyond the standard DFT description, we considered first the pseudopotential self-interaction correction (pSIC) within the Filippetti and Spaldin scheme \cite{filippetti-03,filippetti-09}, as implemented in the QE code \cite{forces}. According to this scheme, the exchange-correlation functional is modified by a set of orbital occupation numbers calculated self-consistently. pSIC has been successfully used to correct some of the failures of the ``standard'' DFT and has the advantage to require relatively modest computational power \cite{filippetti-09}.
Next, many-body effects within the GW approximation \cite{Hedin-65} were taken into account. GW corrections were calculated using the YAMBO code \cite{Marini-09} in the non self-consistent G$_0$W$_0$ scheme on top of the SOC-DFT calculations. 
The frequency-dependent electronic screening was calculated within the plasmon-pole 
approximation \cite{Godby-89}. In order to achieve a convergence of 
0.01~eV in the G$_0$W$_0$ calculations, $k-$point grids of $2\times2\times2$ were 
used for all the supercells. A cutoff of 100~Ry was used for the exchange self-energy. The dielectric matrix was constructed summing up to 1300 unoccupied bands using a cutoff of 5~Ry. Up to 900 unoccupied bands were taken for the G$_0$W$_0$ summation. 

\subsection{Geometric configurations} 
\label{tech2}
To study the effect of the molecular orientations on the electronic structure of MAPbI$_3$ we chose a supercell 
built up from 2$\times$2$\times$2 MAPbI$_3$ unit cells, containing 96 atoms. This supercell is too small to mimic a real disordered crystal, but large enough to show the impact  of the MA orientation on the band structure of the system. Besides, it is consistent with the compatibility between the $C_{3v}$ and $O_h$ symmetries of the MA molecule and the perovskite structure, as well as with the structural coherence length of about 14 \AA\ for MAPbI$_3$ \cite{Choi-14}.
Amongst the large variety of orientations of the MA molecule, 
including disordered arrangements, we chose twenty-four molecular patterns, summarized in Table~\ref{configurations}; to ease the interpretation of this table, Figure~\ref{F2} outlines the atomic arrangement for cases 1 (ordered, Fig.~\ref{F2a}) and 15 (disordered, Fig.~\ref{F2b}).
The first three configurations in Table~\ref{configurations} are homogeneous, with ordered molecules 
oriented along [111], [110] and [100] 
directions, respectively. The subsequent configurations mimicked a ``simulated disorder''. In these, 
MA molecules in adjacent cells were randomly located following some high-symmetry directions, 
namely $\langle 100 \rangle$, $\langle 110 \rangle$ and $\langle 111 \rangle$. 
The cell parameter was taken as its experimental value 6.33~\AA\ \cite{Poglitsch-87}.

Two sets of DFT calculations were carried out. The first set was performed 
at fixed cubic symmetry and $a = 6.33$ \AA, with the atomic positions able 
to change to optimize the Hellmann-Feynman forces between atoms. 
For the second set, both the supercell geometry (lattice parameters and angles) 
and the atomic positions were allowed to change until the pressures over 
the cells were less than 5 bar. In both cases, the atomic positions were optimized 
using the Broyden-Fletcher-Goldfarb-Shanno algorithm until the net force over 
the cell decreased below $7\cdot10^{-3}$~eV$\cdot$\AA$^{-1}$ \cite{Fletcher-87}.

\section{Results and discussion}
\label{results}
\subsection{Formation energies and cell distortions}
\label{dft1}
The formation energies $E_f$ relative to the configuration
with homogeneous molecular orientation along [111] (Table~\ref{configurations}, case 1), taken as reference,
are collected for all the considered geometries in Table~\ref{T1}. 
To examine the effect of the molecular orientations and cage distortions on the stability and the bandgap of the perovskite, we first relaxed the atomic positions keeping fixed the lattice parameter. Starting from the initial configurations schematized in Table~\ref{configurations}, we did not observe substantial deviations of the initial MA orientations upon relaxation.
Next, we allowed the crystal lattice vectors to relax. After relaxation, we observed severe distortions in some cells, while other remained cubic. The molecular configurations maintained the initial orientations for all cases except the structure number 11, which will be discussed below. 

The relative formation energies for the cubic cells vary between -0.2 and 0.7~eV.
This indicates that some of the disordered arrangements (like cases 8, 9, 12 and,
especially 22) are more stable than the ordered one with MA molecules aligned 
along [111]. There are several disordered arrangements 
(such as cases 16, 19 and 21) which, being less stable than the reference configuration, 
have quite similar formation energies. These results suggest that, from whatever initial molecular arrangement, 
the system may evolve under working conditions towards many disordered configurations having similar energy. 

When the cell is allowed to relax, the most stable configuration corresponds to case 1, with case 17 being the most unstable one -formation energy around 1~eV higher than that for case 1. For non-cubic cells we have then the following scenario: if all the molecules are oriented along the [111] direction, the system is in its most stable state and it is unlikely to evolve to any other configuration. On the contrary, if the molecules have arbitrary orientations, then the system can evolve to configurations (disordered in general) which have similar energies. Both relaxations (with fixed or variable cell geometries) are compatible with these transitions, which have been identified by molecular dynamics simulations \cite{Hata-16,mattoni-17}. In addition, we remark that transitions among disordered states are consistent with the symmetry restrictions and agree with experimental findings  \cite{Kojima-09,Baikie-13,Wasylishen-85,Leguy-16b}.

\begin{table}[ht]
\begin{center}
\caption{DFT band gaps $\varepsilon_{gap}$ (in~eV) and 
formation energies E$_f$ (in meV) 
relative to the homogeneous orientation along [111] direction (case 1 of Fig.~\ref{F2}) for fixed cubic cell (columns 2 and 3) and relaxed lattice vectors (columns 4 and 5). 
The atomic positions were allowed to relax in both cases. 
In the last two columns the quasi-gaps $\varepsilon_{q-gap}$ (in~eV) found in the conduction bands of the PbI$_3$ frames, when are distorted like the corresponding cases with MA, are showed.} 
\vspace{1mm}
\begin{tabular}{crcrccc} 
\hline  
 &  \multicolumn{2}{c}{cubic cell} &
    \multicolumn{2}{c}{relaxed cell} &
    \multicolumn{2}{c}{PbI$_3$ frame} \\ 
     &   \multicolumn{2}{c}{relaxed atoms} & 
         \multicolumn{2}{c}{relaxed atoms} & 
         \multicolumn{2}{c}{fixed distortions} \\ 
Case & $E_f^{cubic}$ & $\;\;$ $\varepsilon_{gap}^{cubic}$ & $E_f^{cell}$ & 
   $\;\;$ $\varepsilon_{gap}^{cell}$ & 
   $\;\;$ $\varepsilon_{q-gap}^{cubic}$ &  $\varepsilon_{q-gap}^{cell}$ \\ [0.1cm]
 \hline
 1 &   0 &  1.71 &  0  & 1.58 & 1.70  & 1.60 \\
 2 &  -3 &  1.63 & 145 & 1.68 & 1.68  & 1.69 \\
 3 &  -5 &  1.73 &  89 & 1.72 & 1.68  & 1.75 \\  
 4 &  25 &  1.09 & 624 & 1.25 & 1.43 & 1.54 \\  
 5 & -22 &  1.62 & 211 & 1.54 & 1.67  & 1.61 \\
 6 &  29 &  1.33 & 269 & 1.56 & 1.45  & 1.61 \\   
 7 & -23 &  1.56 &  61 & 1.69 & 1.56  & 1.72 \\  
 8 & -54 &  1.72 & 247 & 1.69 & 1.55  & 1.73 \\
 9 & -81 &  1.58 & 874 & 1.42 & 1.67  & 1.63 \\
10 &  -1 &  1.61 & 128 & 1.74 & 1.61  & 1.73 \\
11 &  70 &  2.18 &  68 & 1.69 & 2.32  & 1.69 \\
12 & -59 &  1.62 & 390 & 1.66 & 1.64  & 1.71 \\
13 & 551 &  1.07 & 837 & 1.29 & 1.40  & 1.55 \\
14 &  75 &  1.42 & 535 & 1.49 & 1.56  & 1.64 \\
15 & -25 &  1.60 & 470 & 1.65 & 1.63  & 1.71  \\
16 & 368 &  1.43 & 770 & 1.51 & 1.52  & 1.54 \\
17 & 699 &  1.21 & 1026 & 1.33 & 1.40  & 1.54 \\
18 & 503 &  1.36 & 845 & 1.49 & 1.41  & 1.54 \\
19 & 348 &  1.39 & 792 & 1.36 & 1.57  & 1.54 \\
20 &  24 &  1.51 & 387 & 1.58 & 1.55  & 1.62 \\
21 & 309 &  1.38 & 783 & 1.36 & 1.62  & 1.66 \\
22 & -201 & 1.65 & 261 & 1.67 & 1.65  & 1.63 \\
23 & 112 &  1.58 & 326 & 1.63 & 1.61  & 1.71 \\
24 & -13 &  1.45 & 454 & 1.49 & 1.62  & 1.68  \\
   \hline
\end{tabular}
\label{T1}
\end{center}
\end{table}

The analysis of the lattice parameters of the several configurations obtained by cell relaxation 
shows that, for ordered molecular arrangements, the cells get 
distorted along the molecular-axis direction, as expected. 
Thus, for direction [110] (case 2) 
one has $a=b=12.91$~\AA$\; $and $c=12.7\;$\AA, while for direction [100] (case 3) 
it is $a=12.97$ \AA$\; $and $b=c=12.84\;$\AA. We also notice that, for these ordered configurations, the cell axes tend to be non orthogonal; actually, the off-diagonal 
components of the metric tensor are non negligible, around 0.25~\AA$\;$in the plane of 
the MA axis. In contrast, for direction [111] (case 1) 
the cell is cubic with $a=b=c=12.90$ \AA.$\; $These results suggest that the cell distortions could be caused by some degree of electrostatic interactions 
between the C or N atoms of the MA molecule and the inorganic frame. They also point out that highly homogeneous molecular patterns relax to lower energy configurations deviating from the cubic structure. 

On the contrary, configurations with antiparallel C-N molecular axes 
cause the lattice vector parallel to the molecule axis to decrease. 
For instance, the cell parameters for case 9 \--which is a $XY$-checkerboard 
of molecular orientations along [001]\-- are $a=b=12.86$ \AA$\;$ and $c=12.82\;$\AA; 
for case 10 \--which is a $XYZ$-checkerboard along [001]\-- they are 
$a=b=12.85\;$\AA$\;$ and $c=12.76\;$\AA.  
For the most disordered configurations, for which not all the molecular axes are oriented parallel or antiparallel to each other, more complex distortions are observed upon relaxation. 
Very recently, Quarti \textit{et al.} used Car-Parrinello molecular dynamics to show that the cubic perovskite structure of MAPbI$_3$ 
exhibits large distortions at the sub-picosecond time-scale \cite{quarti-2016}. This finding, which explains part of the features of the tetragonal-cubic phase transition, 
shows the importance to consider several structures and MA arrangements in order to understand electronic and optical properties of MAPbI$_3$.

The change of the lattice parameters cause the characteristic PbI$_6$ octahedra of the inorganic cage to distort as well. This effect has been widely described in the literature, so that it will not be described further herein. Here we will just mention that the distortions of the inorganic cage obtained in this work for the most symmetrical configurations agree with the results reported elsewhere \cite{mosconi-13,motta-15,quarti-2016,Stoumpos-13,Beecher-16}.

\subsection{Electronic band gaps}
\label{dft2}

\begin{figure}
\centering
\includegraphics[scale=0.30]{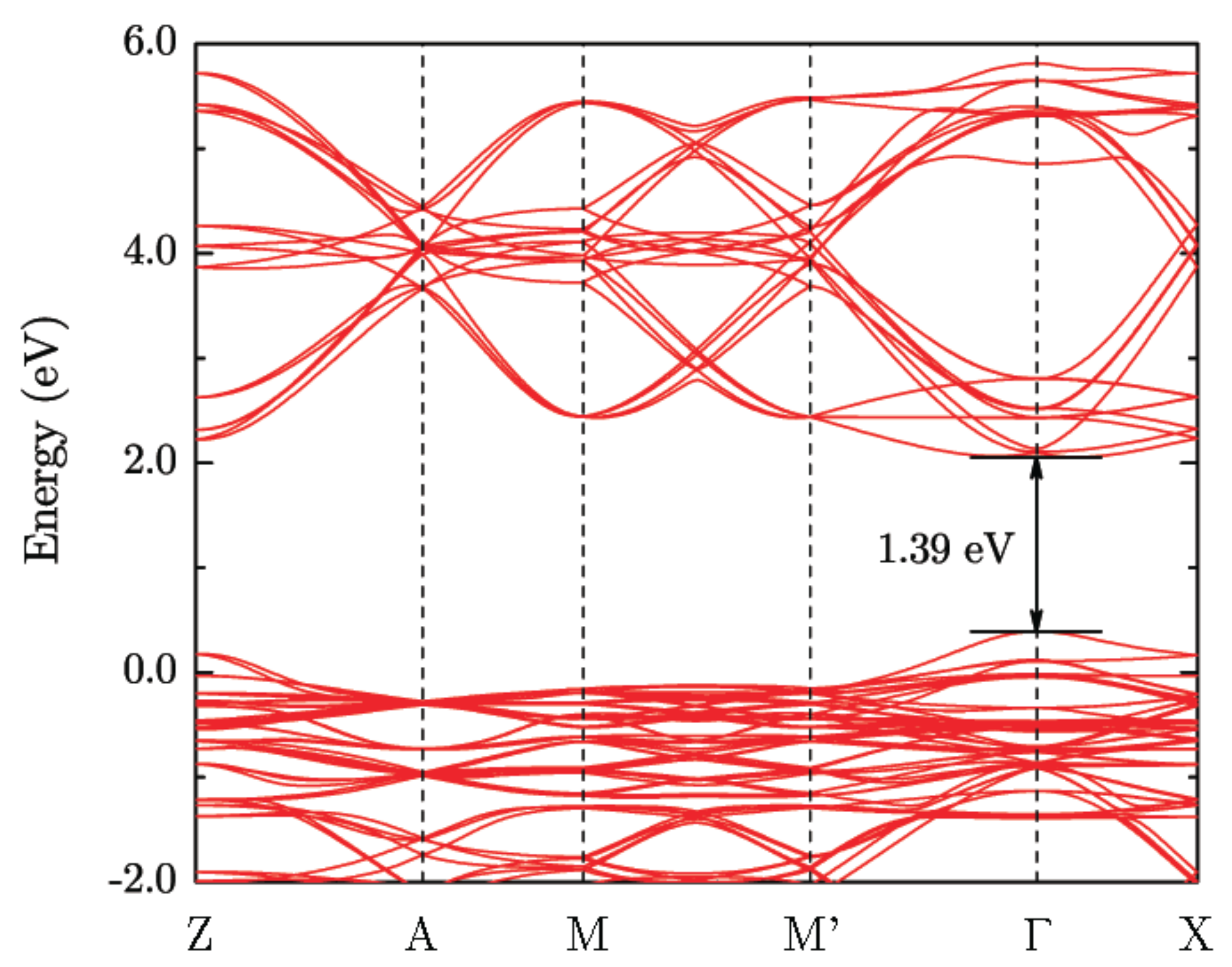}
\caption{Band structure of the perfect non-distorted cubic PbI$_3$ crystal 
without MA, calculated within standard DFT for the 2$\times$2$\times$2 supercell.
The Fermi level (horizontal black line) is set at 0~eV.}
\label{F3}
\end{figure}

The calculated band gaps for all the studied configurations, for both sets of relaxations, are reported in Table~\ref{T1} as well.
PbI$_3$ is a poor metal with a ``quasi-gap'' in the conduction band manifold, as shown for the cubic, 
non-distorted supercell in Fig.~\ref{F3}. When the MA molecule is incorporated, the Fermi level rises up, so that the ``quasi-gap" of PbI$_3$ becomes 
the true band gap of MAPbI$_3$.
Therefore, for a comparison with the band gaps of the whole crystal, we also report the ``quasi-gap'' in distorted PbI$_3$ frames (that is 
the ``quasi-gap'' calculated for the PbI$_3$ cage without the MA molecules, but with Pb and I at the positions that they occupy in the 
corresponding cases) calculated under the same conditions: keeping the cubic cell symmetry ($\varepsilon_{q-gap}^{cubic}$) and allowing 
the cubic cell to relax ($\varepsilon_{q-gap}^{cell}$). 

The calculated band gaps for the cubic cells vary by more than 1~eV, between 1.07 and 2.18~eV.
Interestingly, the largest gap corresponds to case 11, for which the most severe distortions of the cage are observed. In this case, the final configurations presents a Pb - I - Pb angle as low as around 167$^{\circ}$ and 
the MA molecules no longer align along $\langle 111 \rangle$ directions. 
We will present below the distortions for case 11, to be compared to those for cases 1 and 4.

By contrast, the atomic displacements are modest for 
the configurations with the narrowest gaps. 
When the cells are allowed to relax, the gaps range between 1.25 and 1.74~eV, the largest one corresponding now to case 10. 
We notice that the ``quasi-gaps'' of the PbI$_3$ frame follow roughly the decreasing/increasing trend and the size order of the corresponding MAPbI$_3$ band gaps in both cases.
Table~2 collects
selected recently calculated band gaps for the MAPbI$_3$ system, while in Table~\ref{T3} experimental values are shown for comparison.

We observed that band gaps calculated within DFT seem to increase as the symmetry of the unit cell decreases \cite{Baikie-13}; this effect 
is common to other hybrid perovskite systems \cite{Borriello-08}, and has been experimentally confirmed for the orthorhombic to tetragonal 
phase transition in MAPbI$_3$ \cite{Xing-14}. On the other hand, the band gaps are sensitive not only to the orientation of the MA 
molecules, but also to the size of the unit cell for a given crystal system. Thus, Amat \textit{et al.} find a small shift from 1.34 to 
1.30~eV for a cubic cell when the cell volume decreases from 254 \AA$^3$ to 251 \AA$^3$; the band gap variation is more important 
(from 1.66 to 1.56~eV) for a volume change from 248 \AA$^3$ to 244 \AA$^3$ in tetragonal cells \cite{Amat-14}. Our results in 
Table~\ref{T1} are in fair agreement with those reported in Table~\ref{T2}, and confirm this 
trend for both the fixed and the variable cell geometries with some exceptions. For the fixed cubic cell geometries, case 11 yields a high 
gap of 2.18~eV, whereas cases 4 and 13 yield reduced gaps of around 1.1~eV; for the relaxed cells, cases 4 and 13 yield gap values below 1.3~eV.
In other words, the DFT results suggest that realistic disordered molecular arrangements within the  MAPbI$_3$ structure, which are allowed by energetic 
considerations, may give rise to gaps significantly lower than those previously reported considering oriented molecules. 

\begin{table}
\renewcommand{\arraystretch}{0.9}
\begin{center}
\caption{Selected theoretical (DFT and beyond) for the band gap of the MAPbI$_3$ system. DFT-vdW: DFT calculation with van der Waals corrections; QSGW: Quasiparticle self-consistent GW \cite{Schilfgaarde-06}; scGW$_0$: self-consistent-eigenvalue GW.}
\begin{tabular}{>{\centering\arraybackslash}m{3.5cm} >{\centering\arraybackslash}m{2.5cm} >{\centering\arraybackslash}m{2cm} >{\centering\arraybackslash}m{2cm}} 
& & &  \\
Reference & $\varepsilon_g$ (eV) & Geometry & Method \\ 
\hline
\rowcolor{gray!20}
\multicolumn{4}{c}{Theoretical Results}\\
 & 1.30 & Cubic &  \\
Baikie \textit{et al.} \cite{Baikie-13} & 1.43 & Tetragonal & DFT \\ 
 & 1.61 & Orthorhombic & \\ 
\hline
Umebayashi \textit{et al.} \cite{Umebayashi-03} & 1.45 & Cubic & DFT \\
\hline
\multirow{2}{*}{Motta \textit{et al.} \cite{motta-15}} & 1.5 - 1.7 & \multirow{2}{*}{Cubic} & DFT \\ 
 	& 1.42 - 1.61 & & DFT-vdW \\ 
\hline
\multirow{4}{*}{Umari \textit{et al.} \cite{Umari-14}} & 1.68 & \multirow{4}{*}{\centering Tetragonal} & DFT \\ 
	& 0.60 & & SOC-DFT \\ 
    & 2.68 & & GW  \\ 
    & 1.67 & & SOC-GW \\
\hline
 \multirow{5}{*}{Amat \textit{et al.} \cite{Amat-14}}  & 1.30-1.34  &  \multirow{2}{2cm}{\centering Cubic} & DFT  \\ 
  & 0.07 - 0.11 & & SOC-DFT  \\ \cline{2-4}
  & 1.56-1.66 & \multirow{3}{2cm}{\centering Tetragonal} & DFT \\ 
  & 0.57 & & SOC-DFT \\
  & 1.67 & & SOC-GW \\  \cline{2-4}
\hline
\multirow{2}{*}{Zhu \textit{et al.} \cite{Zhu-14}} & 1.57 & Tetragonal & SOC-GW \\
 	& 1.69 & Orthorhombic & SOC-GW\\ \hline
Leguy \textit{et al.} \cite{Leguy-16} & 1.46 - 1.60 & Cubic & SOC-QSGW \\
\hline
\multirow{4}{*}{Scherpelz \textit{et al.} \cite{Scherpelz-16}} & 1.49 & \multirow{4}{*}{\centering Orthorhombic} & DFT \\ 
	& 0.44 & & SOC-DFT \\ 
    & 2.84 & & GW  \\ 
    & 1.51 & & SOC-GW \\
\hline
\multirow{4}{*}{Filip \textit{et al.} \cite{filip-14b}} & 1.50 & \multirow{5}{*}{\centering Orthorhombic} & DFT \\ 
	& 0.58 & & SOC-DFT \\ 
    & 2.55 & & GW  \\ 
    & 1.32 & & SOC-GW \\
    & 1.79 & & SOC-scGW\\
\hline
\end{tabular}
\label{T2}
\end{center}
\end{table}

\begin{table}
\renewcommand{\arraystretch}{0.9}
\begin{center}
\caption*{Continuation of Table~3: 
Selected theoretical (DFT and beyond) for the band gap of the MAPbI$_3$ system. DFT-vdW: DFT calculation with van der Waals corrections; QSGW: Quasiparticle self-consistent GW \cite{Schilfgaarde-06}; scGW$_0$: self-consistent-eigenvalue GW.}
\begin{tabular}{>{\centering\arraybackslash}m{3.5cm} >{\centering\arraybackslash}m{2.5cm} >{\centering\arraybackslash}m{2cm} >{\centering\arraybackslash}m{2cm}}
& & &  \\
Reference & $\varepsilon_g$ (eV) & Geometry & Method \\ 
\hline
\rowcolor{gray!20}
\multicolumn{4}{c}{Theoretical Results}\\
\multirow{4}{*}{Ahmed \textit{et al.} \cite{Ahmed-15}} & 1.51 & \multirow{4}{*}{Cubic} & DFT \\ 
 	& 0.46 & & SOC-DFT \\ 
    & 2.53 & & GW \\ 
    & 1.48 & & SOC-GW\\ 
\hline
\multirow{3}{*}{Quarti \textit{et al.} \cite{quarti-2016}} & 1.16/1.28  & Cubic & \multirow{3}{*}{SOC-GW} \\
 	& 1.67 & Tetragonal & \\ 
    & 1.81 & Orthorhombic & \\
    \hline
    \multirow{5}{*}{Brivio \textit{et al.} \cite{Brivio-14}} & 1.38 & \multirow{5}{*}{Cubic} & \multirow{2}{*}{DFT} \\
	& 1.46 & & \\
    & 0.53 & & SOC-DFT \\
    & 1.27 & & SOC-GW  \\
    & 1.67 & & SOC-QSGW\\
\hline
Bokdam \textit{et al.}~\cite{Bokdam-16} & 1.67 & Cubic & SOC-scGW$_0$\\
\hline
\end{tabular}
\end{center}
\end{table}

\begin{table}
\renewcommand{\arraystretch}{1.2}
\begin{center}
\caption{Selected experimental results  for the band gap of the MAPbI$_3$ system. IPCE: Incident photon-to-current quantum conversion efficiency; PES: Photoemission Spectroscopy; OS: Optical Spectroscopy; PL: Photoluminiscence; UTA: Ultrafast Transient Absorption.}
\begin{tabular}{>{\centering\arraybackslash}m{4cm} >{\centering\arraybackslash}m{2.5cm} >{\centering\arraybackslash}m{2cm} >{\centering\arraybackslash}m{2cm}}
& & &  \\
Reference & $\varepsilon_g$ (eV) & Geometry & Method \\ 
\hline
\rowcolor{gray!20}
\multicolumn{4}{c}{Experimental Results}\\
Baikie \textit{et al.} \cite{Baikie-13} & 1.51 & \multirow{4}{*}{Tetragonal} & OS \\ \cline{1-2}\cline{4-4}
Kojima \textit{et al.} \cite{Kojima-09} & 1.55 &  & IPCE \\ \cline{1-2}\cline{4-4}
Stoumpos \textit{et al.} \cite{Stoumpos-13} & 1.52 &  & OS \\ \cline{1-2}\cline{4-4}
Stamplecoskie \textit{et al.} \cite{stamplecoskie-15} & 1.63  & & OS \\ \hline
Schulz \textit{et al.} \cite{Schulz-14} & 1.70 & \multirow{2}{2cm}{\centering Grown on TiO$_2$} & PES \\ \cline{1-2}\cline{4-4}
Yamada \textit{et al.} \cite{Yamada-14} & 1.61 & & PL \\ \hline
Manser \textit{et al.} \cite{Manser-14} & 1.63 & \multirow{4}{2cm}{\centering Grown on Al$_2$O$_3$} & OS \\ \cline{1-2}\cline{4-4}
Grancini \textit{et al.} \cite{Grancini-15} & 1.67 & & \multirow{3}{*}{UTA} \\ \cline{1-2}
Sheng \textit{et al.} \cite{Sheng-15} & 1.66 & & \\ \cline{1-2}
Price \textit{et al.} \cite{Price-15} & 1.58 & &  \\ \hline
Ishihara \cite{Ishihara-94} & 1.63  & --- & OS \\ 
 \hline
\end{tabular}
\label{T3}
\end{center}
\end{table}

The impact of the molecular orientation on the band gap 
points to important implications regarding the optical performance of MAPbI$_3$, although additional effects (namely the spin-orbit coupling and many-body corrections) remain to be considered. We will discuss this in section \ref{SOC} below.
Important questions are whether the lattice vectors actually relax under experimental conditions, in which grain boundaries exist and molecules rotate fast, and how quickly the heavy inorganic cage can follow the rotation of the MA pattern \cite{Hata-16}. Moreover, in real samples with many molecular-pattern domains, 
both the overall pattern and domain walls can move. This results in a system which is not periodic at all, as has been proved by a direct observation of the dynamic symmetry breaking, with an estimate of the domain size of 1-3 nm, by atomic pair distribution function measurements \--an effect which is non observable by Bragg diffraction \cite{Beecher-16}.
Finally, we should take into account that the device at work is in the external electric field caused by the photovoltaic charge on the electrodes, which tends to align the molecules parallel to it. Thus, the actual molecular arrangement would arise from the balance between the ordering effect of the field and the disordering effect of temperature.

\subsection{Origins of the band-gap variation}
\label{dft3}

\begin{figure*}
\centering
\includegraphics[width=1.1\textwidth]{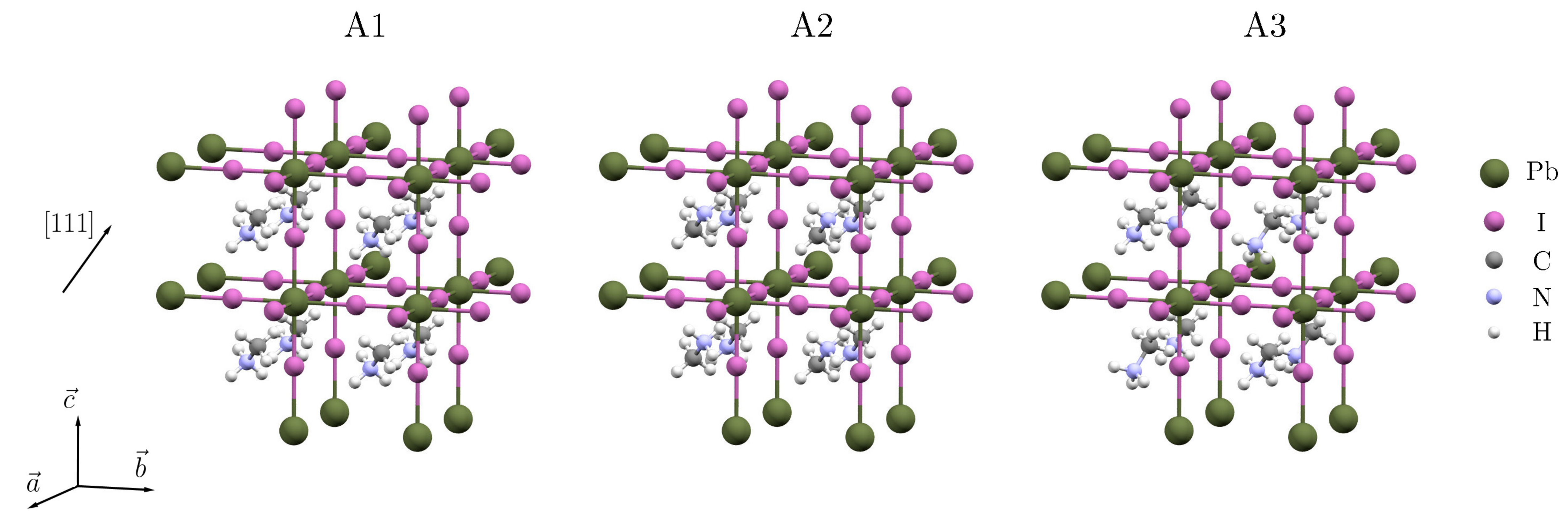}
\caption{Molecular arrangement for configurations A1, A2 and A3.}
\label{F4}
\end{figure*}

In this section we analyze in detail the band structure for three configurations in order to rationalize the cell distortions in terms 
of interactions between the C or N atoms of MA and the Pb or I atoms of the inorganic cage.
We chose three cases for simplicity: i) the homogeneous pattern along [111] direction, case 1 in Table~\ref{configurations}, which has the lowest formation energy when the cell is relaxed;
ii)  the configuration with antiparallel C-N ends, case 4 in 
Table~\ref{configurations}, which has the smallest band gap; and iii) the "anti-diagonal" zigzag pattern, case 11 in Table~\ref{configurations}, which has the largest band gap in the cubic cell, as well as the largest atomic displacements upon relaxation. Figure~\ref{F4} schematizes these molecular arrangements, that will be hereafter referred to as A1, A2 and A3, respectively.

\begin{figure}[!ht]
	\centering
	\subfloat[A1]{\label{subfig:bands_a1}\includegraphics[width=0.52\textwidth]{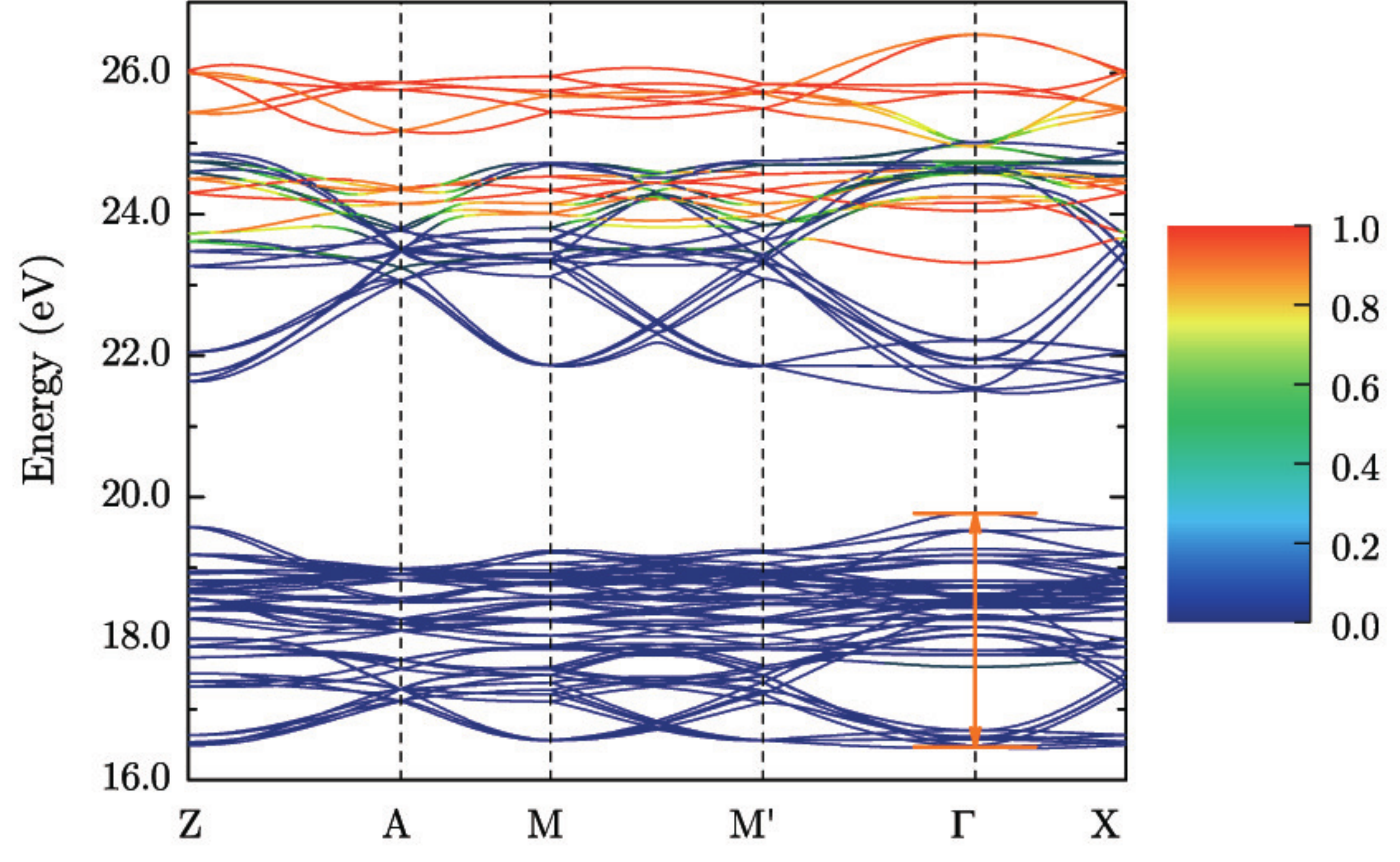}} \\
	\subfloat[A2]{\label{subfig:bands_a2}\includegraphics[width=0.52\textwidth]{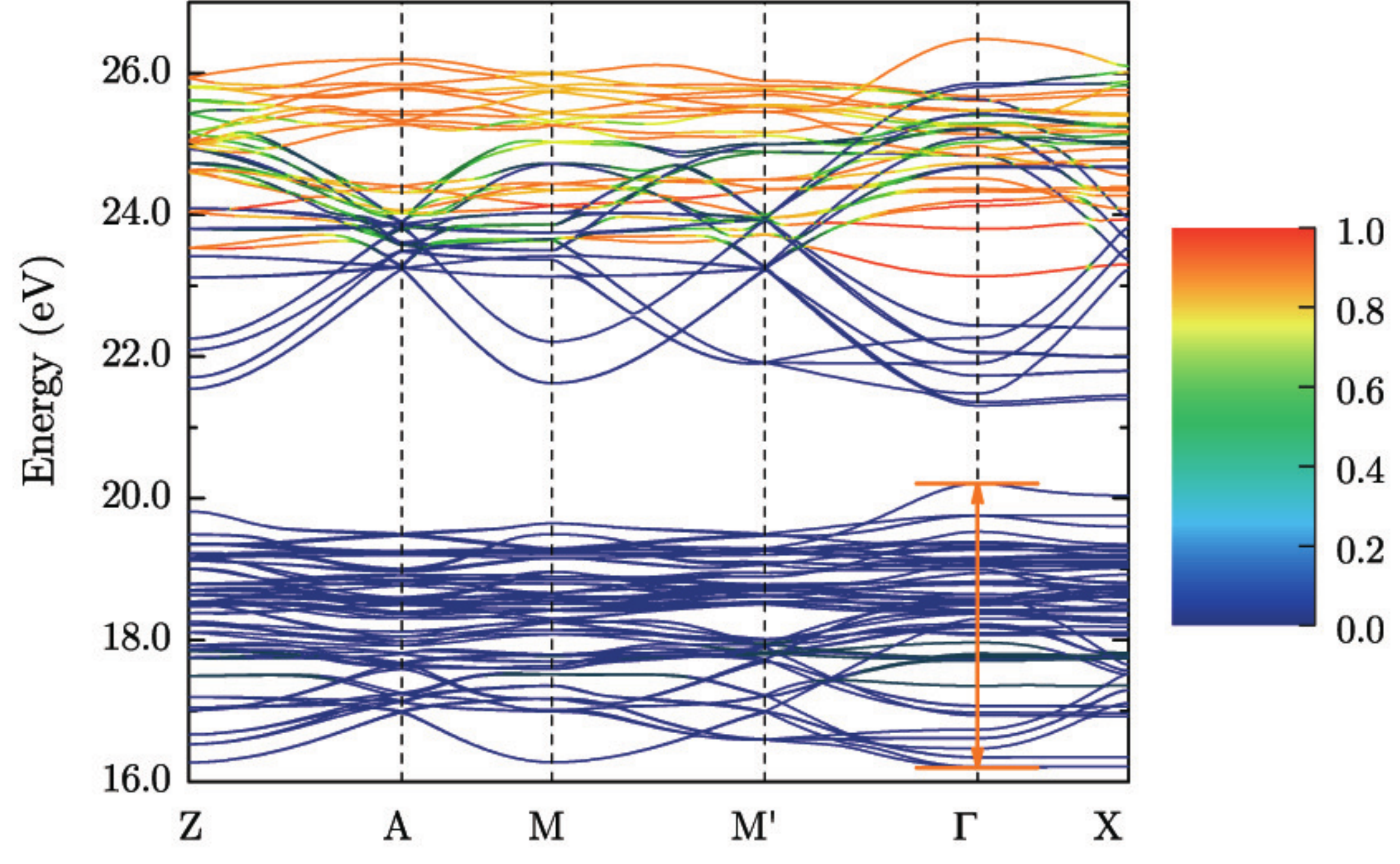}} \\
	\subfloat[A3]{\label{subfig:bands_a3}\includegraphics[width=0.52\textwidth]{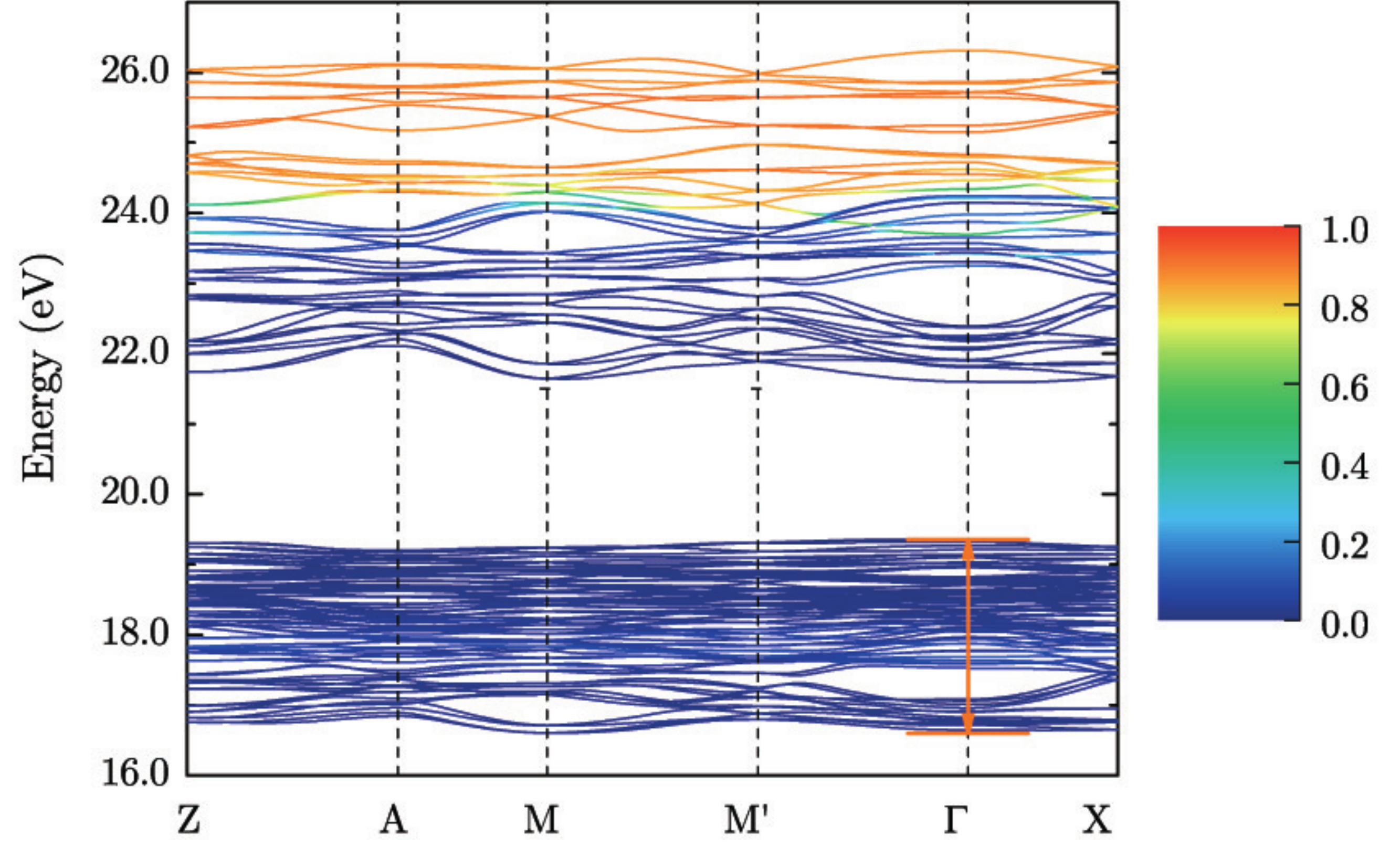}}
	\caption{Band structure near the corresponding gaps for A1, A2 and A3 projected onto MA states. In these figures, the reference zero energy is set at the Pb-$6s$ state (not shown).}
	\label{F5}
\end{figure}

\begin{figure}
\centering
\includegraphics[width=\textwidth]{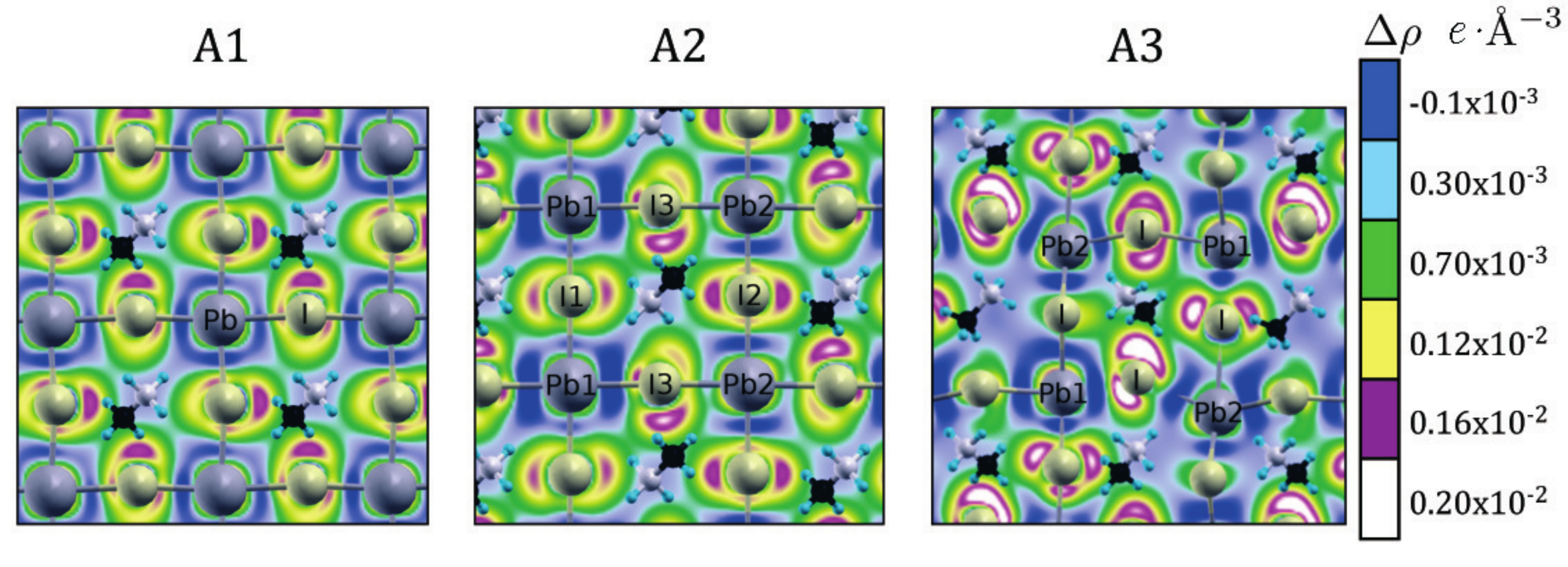}
\caption{Differential charge density maps of MAPbI$_3$ across 
the PbI$_3$ frame in the $XZ$ 
plane for cases A1, A2 and A3 with respect to the corresponding 
PbI$_3$ frames in absence of the MA molecules. 
The C and N atoms of MA are plotted in white and black, respectively.}
\label{F6}
\end{figure}

Figure~\ref{F5} shows the valence 
and conduction band manifolds of the A1, A2 and A3 configurations, 
calculated within DFT, projected onto MA-localized Wannier functions. 
The color scale gives the coefficient of the MA-Wannier function in 
the total composition of the Bloch states. In all cases, deep flat bands 
localized on Pb-$6s$ have almost identical energies and are used 
as reference (zero energy in the bands plot, not shown).
From Fig.~\ref{F5} one first notices that the topmost valence and
bottommost conduction bands have no molecular character; the MA states give rise,
on the contrary, to conduction bands well above the gap. It is also noticeable that
the bands topology varies much with the degree of disorder; in particular, substantially
flatter bands are observed for the A3 case, the most disordered one. In addition,
according to Fig.~\ref{F5}, 
the differences found in the gap size for the three cases can be ascribed 
mainly to the position of the valence band top (VBT), denoted with a
horizontal short line. It varies to within 
almost 1.0~eV, while the conduction band minimum (CBM) is rather constant 
(variation of about 0.2~eV). Moreover, differences of about 0.4~eV  
are also found for the position of the bottom of the valence-band manifold, 
in opposite direction to the VBT change. As a result, the valence band width 
(shown as a doble arrow in Figs.~\ref{F5}) 
is the largest for the case with the smallest gap (A2) and the smallest for the case 
with the largest band gap (A3). 

In order to investigate the role of the interactions of the MA molecules 
with the inorganic cage, we show in Figure~\ref{F6} the color maps 
representing the differences between the charge density of the whole 
system and that without MA molecules, but with the same inorganic cage. 
From these plots, one can see that the N-end induces negative charge near I, 
and positive charge near Pb. In the end, this suggests that electrostatic effects 
caused by the molecular orientations are responsible for the structural 
distortions and gap variations.

\begin{figure*}[!ht]
	\centering
	\subfloat[A1]{\label{subfig:coord_a1}\includegraphics[scale=0.12]{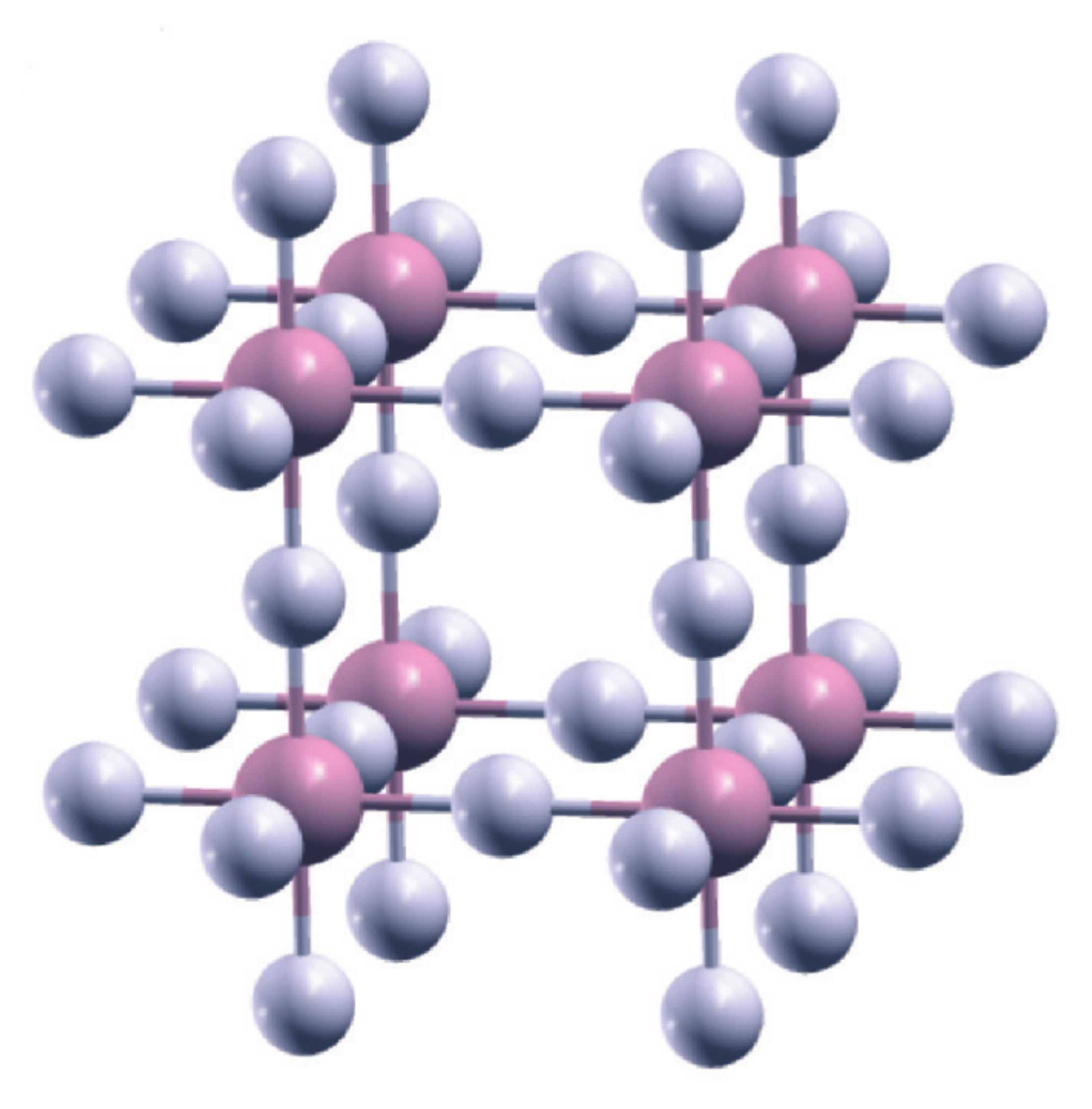}} \hspace{0.5cm}
	\subfloat[A2]{\label{subfig:coord_a2}\includegraphics[scale=0.12]{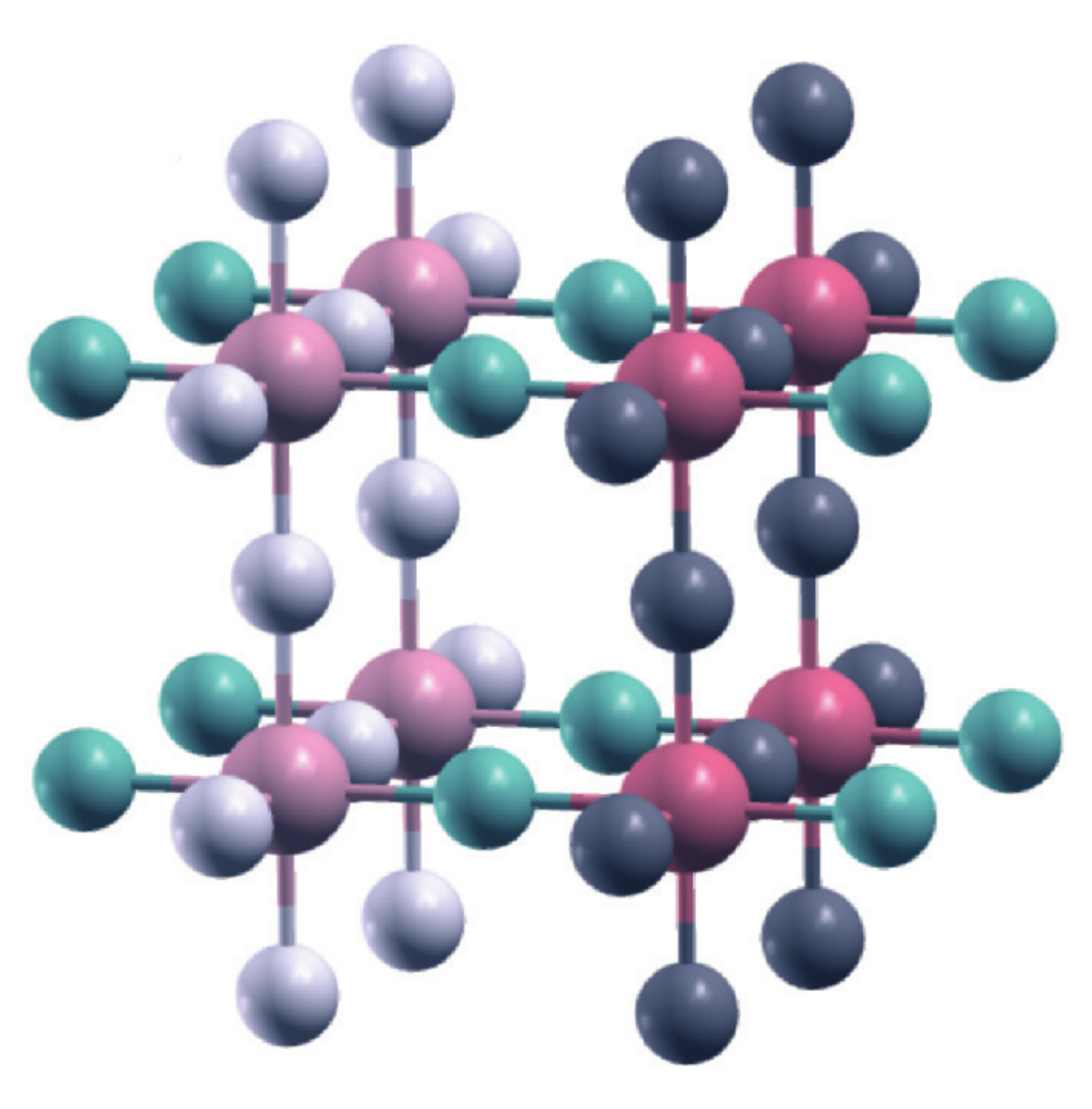}} \hspace{0.5cm}
	\subfloat[A3]{\label{subfig:coord_a3}\includegraphics[scale=0.12]{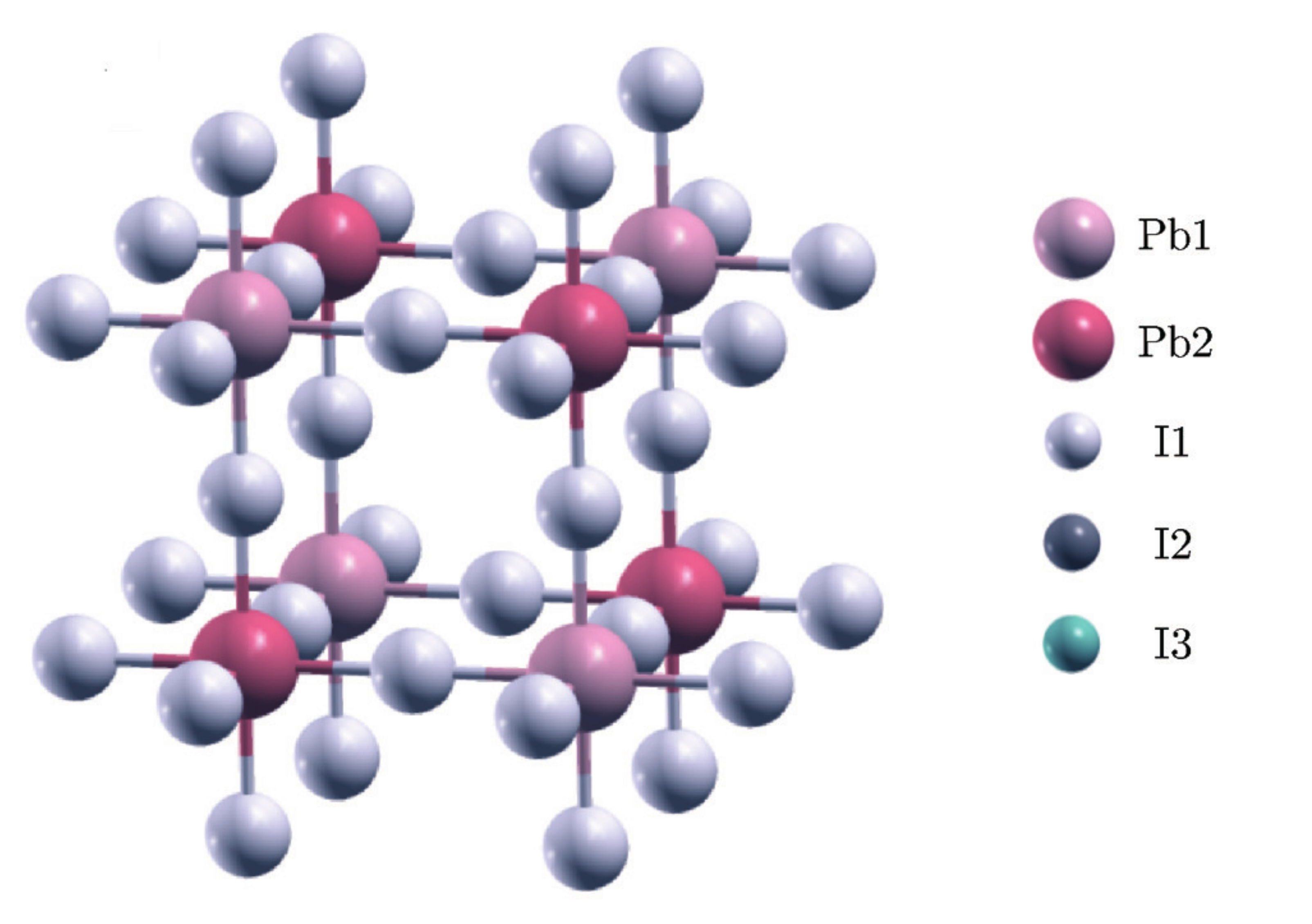}}\\
	\subfloat[A1]{\label{subfig:projections_a1}\includegraphics[width=0.6\textwidth]{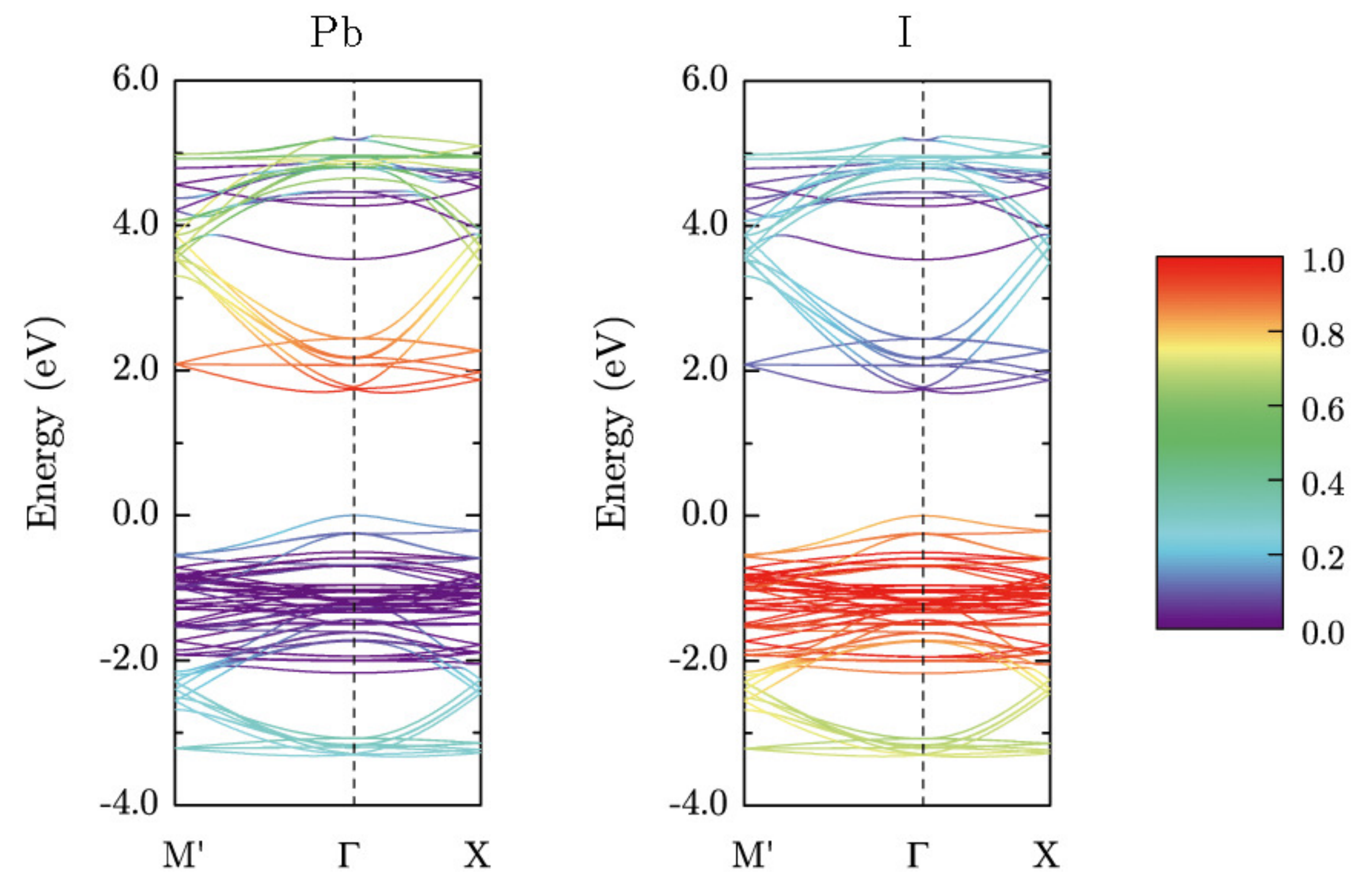}} \\
	\subfloat[A2]{\label{subfig:projections_a2}\includegraphics[width=0.85\textwidth]{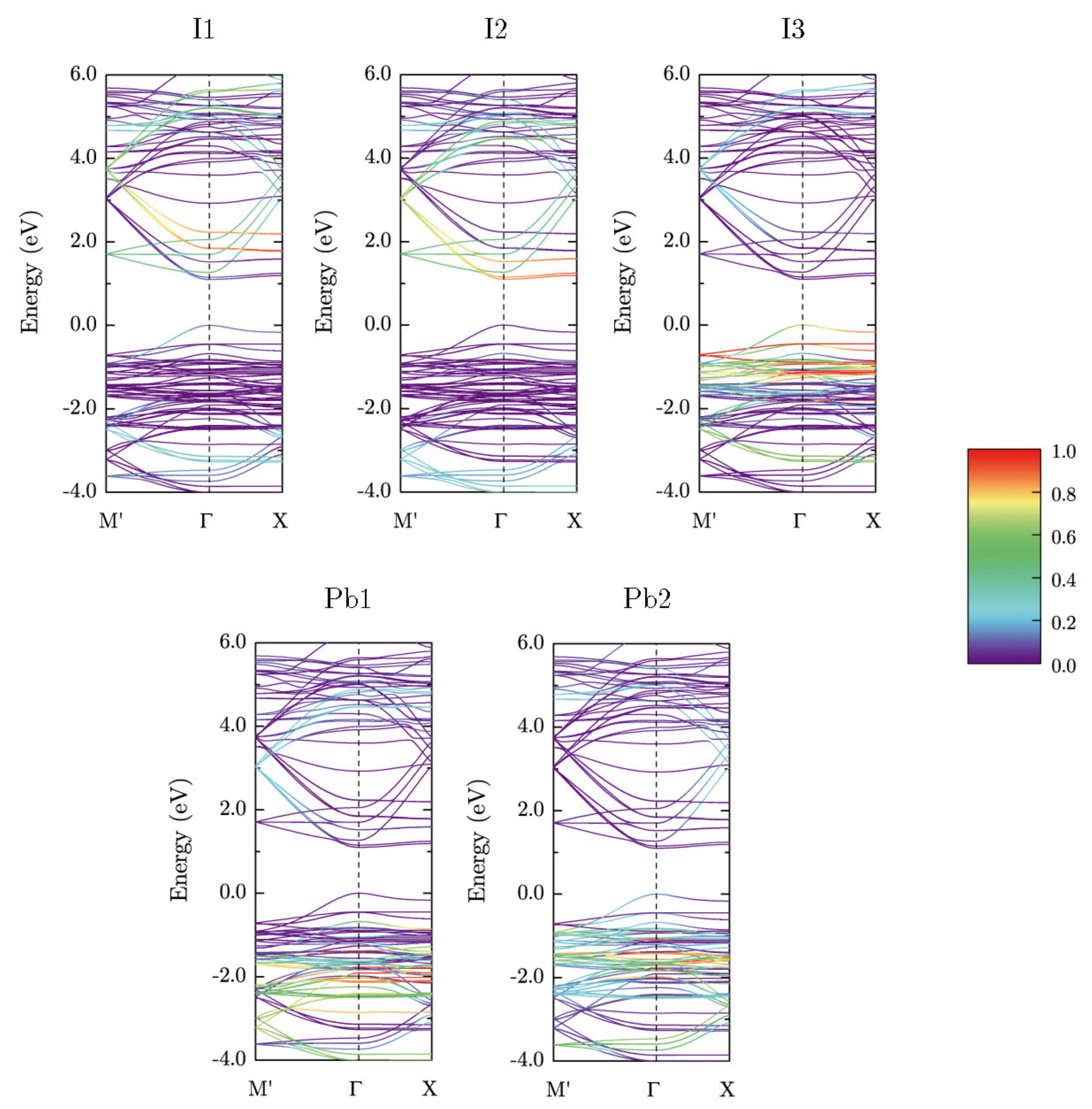}} 
\end{figure*}

\begin{figure}[!ht]
	\centering
	\subfloat[A3]{\label{subfig:projections_a3}\includegraphics[width=0.82\textwidth]{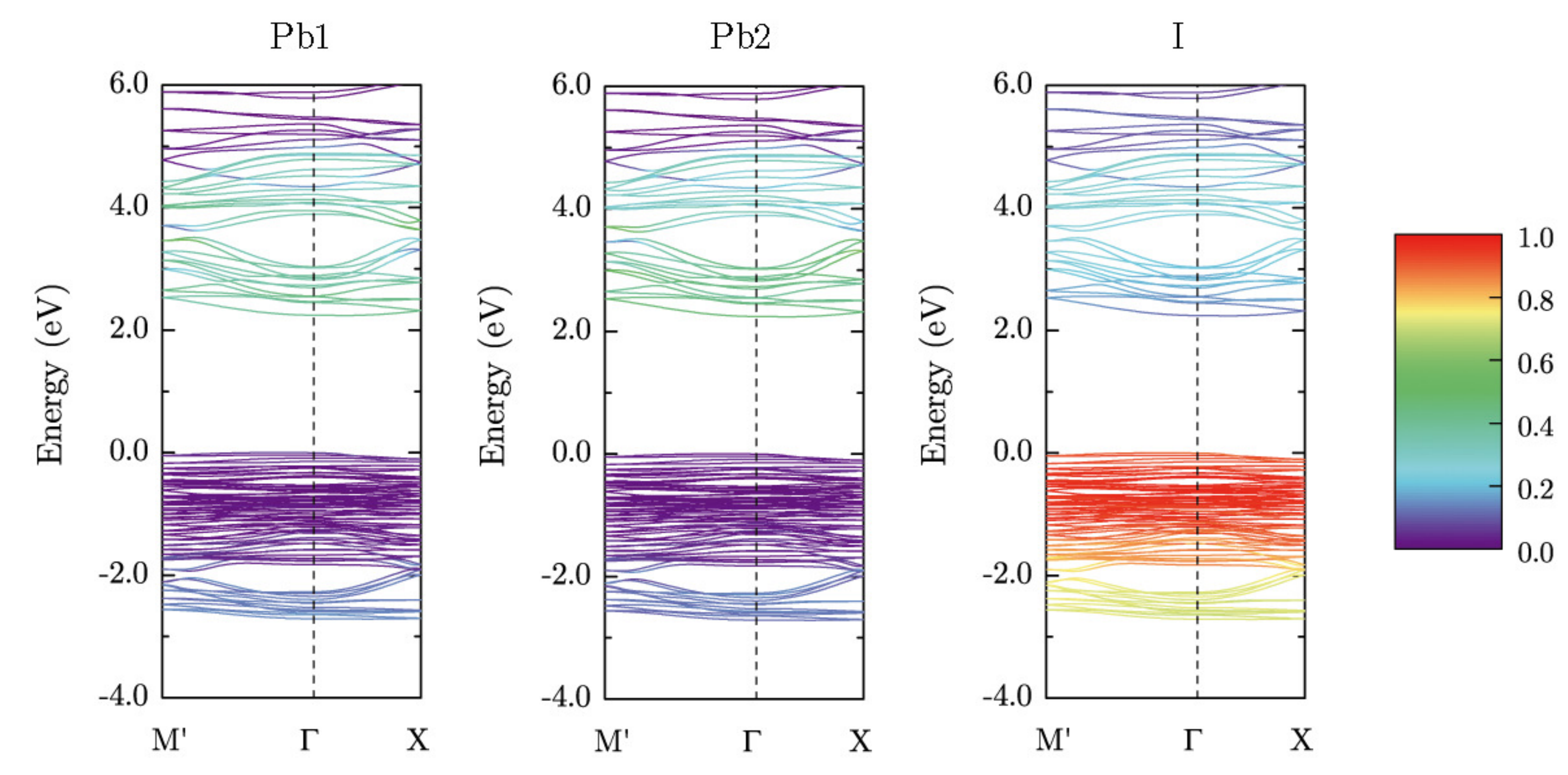}}\\
	\caption{Top: Ball and stick representations of the inorganic cages for structures A1, A2 and A3: nonequivalent atoms are drawn in different color. Bottom: Band structures for the same configurations projected onto the symmetry nonequivalent Pb and I atoms. The Fermi energy is set at 0~eV.}
	\label{F7}
\end{figure}

To further investigate the differences in the electronic structure for cases A1, A2 and A3 we may distinguish several
Pb and I atoms depending on their local symmetries. Indeed, since 
each MA pattern is characterized by the number and orientation 
of C- and N-ends around the nearest Pb and I atoms, these frame centers 
can be divided into groups depending on their local ``coordination'' number. 
The symmetry inequivalent Pb and I atoms for cases A1, A2 and 
A3 are outlined by different colors in Figs.~\ref{subfig:coord_a1}, \ref{subfig:coord_a2} and
\ref{subfig:coord_a3}, respectively. 
For the A2 pattern, one can find two non-equivalent Pb and three I atoms, 
whereas in the homogeneous case A1, all Pb and I centers are equivalent 
by symmetry. For the lowest symmetry case A3, we find two non-equivalent 
groups of Pb atoms. Figs.~\ref{subfig:projections_a1}, \ref{subfig:projections_a2} and
7f show the projections of the bands onto Wannier functions localized around 
the symmetry inequivalent Pb and I atoms for the cases A1, A2 and A3, respectively.
These results indicate that, in all cases, 
the VBT has mainly I-$5p$ character, whereas the CBM is essentially 
Pb-$6s,6p$ type; this is easily visualized for A1 in Fig.~\ref{F7}. 
In A2, the configuration with the smallest gap, the VBT arises from the 
three symmetry non equivalent atoms (namely I-1, I-2 and I-3), 
and its projections onto the corresponding atomic states are different. 
This results in lifting the VBT degeneracy which, in turn, shrinks the band gap.

\subsection{Spin-orbit, many-body and self-interaction correction effects}
\label{SOC}
The 
agreement between experimental band gaps and those calculated within the DFT scheme (discussed in Tables \ref{T1} and 3) is somewhat surprising, since it is well known that DFT inherently 
underestimates band gaps. For MAPbI$_3$, it has been argued that such a good agreement is fortuitously due to error cancellations when
spin-orbit coupling (SOC) corrections are neglected. 
Several authors have recently showed the importance to include SOC effects due to the presence of the heavy Pb atoms, with the consequence of a drastic band gap reduction~\cite{Umari-14,Zhu-14,Amat-14,Brivio-14,Ahmed-15,Leguy-16}.

We have included SOC effects by using
fully-relativistic pseudopotentials for a subset of 12 configurations chosen from those in Table~\ref{configurations}. 
To ease the analysis, the calculations were carried out for fixed cubic symmetry (that is, neglecting the cell distortions). The 
SOC-DFT band gaps of the 12 selected cases are collected in Table~\ref{T4}.
In all cases, the SOC indeed reduces severely the band gaps of MAPbI$_3$ to values much narrower than those experimentally measured, in good agreement with reports by other authors  included in Table~\ref{T1}. It is remarkable that the effect of the SOC is not uniform for all the cases. In essence, the effect of the molecular ordering on the SOC-DFT band gaps is the same as for the DFT without SOC, but the band gap shrinkage depends on the particular case, ranging roughly between 0.8~eV (case 11) and 1.3~eV (case 9). We have then a scenario in which ``standard'' DFT fortuitously yields band gaps which agree 
with experiments, but the required SOC corrections shrink them to unrealistic values. 

It is thus evident that more sophisticated levels of approximation are required 
to account for the experimental findings. To get further insight on the band gaps 
of the studied systems beyond DFT, we have considered two different approaches. 
First is the inclusion of the self-interaction correction (SIC) within DFT via 
pseudo-self-interaction-corrected density functional theory (pSIC) \cite{filippetti-03}. 
Indeed, DFT is affected from self-interaction errors, which are especially 
pronounced in strongly correlated systems containing electrons in the $d$- or 
$f$-atomic shells. pSIC has been showed to provide reliable electronic 
and magnetic properties of pure or doped correlated systems as well~\cite{filippetti-09}.

In the cases considered herein, we have localized valence states from the molecules, suggesting that SIC could play a role. 
pSIC calculations were carried on using the QE code, and the effect of the SIC on the gaps of the different configurations was estimated by considering the correction obtained without the inclusion of spin-orbit effect, because of the unavailability of a scheme that allows inclusion of SOC in pSIC calculations.
The results are  collected in Table~\ref{T4}.

\begin{table}[h]
\begin{center}
\caption{Band gaps in~eV for a subset of configurations taken from Fig.~\ref{F2} in different approximations:
 $\Delta$=pSIC-DFT.}
\begin{tabular}{cm{0.8cm}>{\centering\arraybackslash}m{1.1cm}>{\centering\arraybackslash}m{1.1cm}>{\centering\arraybackslash}m{2.1cm}>{\centering\arraybackslash}m{1.1cm}>{\centering\arraybackslash}m{2.1cm}}  \hline  
Case & DFT & pSIC & $\Delta$ & SOC-DFT & SOC+$\Delta$ & SOC-GW \\[0.2cm]
\hline \\
 1 &  1.71 &  2.25 & 0.54 & 0.69 & 1.23 & 1.20 \\
 2 &  1.63 &  2.22 & 0.59 & 0.68 & 1.27 & 1.12 \\
 3 &  1.73 &  2.17 & 0.44 & 0.65 & 1.09 & 1.29 \\
 4 &  1.09 &  1.57 & 0.48 & 0.06 & 0.54 & 0.42 \\
 5 &  1.62 &  1.84 & 0.22 & 0.43 & 0.65 & 1.01 \\
 7 &  1.56 &  2.11 & 0.55 & 0.52 & 1.07 & 0.87 \\ 
 8 &  1.72 &  1.94 & 0.22 & 0.41 & 0.63 & 1.30 \\
 9 &  1.58 &  1.64 & 0.06 & 0.32 & 0.38 & 1.08 \\
10 &  1.61 &  2.17 & 0.56 & 0.62 & 1.18 & 1.11 \\
11 &  2.18 &  2.19 & 0.01 & 1.42 & 1.43 & 2.33 \\
13 &  1.07 &  1.56 & 0.49 & 0.04 & 0.53 & 0.40 \\
17 &  1.21 &  0.71 & 0.50 & 0.12 & 0.62 & 0.55 \\
\\[-0.2cm]
\hline 
\end{tabular}
\label{T4}
\end{center}
\end{table}

As for the SOC correction, our results show that the effect of the SIC is not uniform for all the cases. It is stronger for ordered cases, namely cases 1 and 2, with parallel orientation; however, the largest gap is exhibited by case 11, as for ``standard'' DFT.
On the contrary, the SIC are weak for anti-parallel
orientations (cases 4, 13 and especially 9). Ultimately, the self-interaction corrections, as the DFT values themselves, result to be 
highly dependent on the interactions between 
the MA molecules and the inorganic frame.
Considering, as a first approximation, the SIC simply added to the SOC-DFT energy gaps (SOC+$\Delta$ in Table~\ref{T4}),  we observe differences in the fundamental gap up to 1~eV depending on the  molecular orientations.

The second level of approximation beyond DFT is MBPT. In order to include many-body effects, we performed an additional series of calculations on selected cases within the GW approximation \cite{Hedin-65}.
GW calculations were done including spin-orbit effects (SOC-GW in Table~\ref{T4}), that is, considering spinorial wave functions starting from the SOC-DFT ground state. The GW approximation was taken in the so-called G$_0$W$_0$ scheme, so that no self-consistency was considered in the solution of the quasi-particle equation.

Last column of Table~\ref{T4} records the G$_0$W$_0$ results for the band gaps. The effect of many-body effects is to widen the gaps (which range now roughly between 0.40 and 2.33~eV) with respect to the SOC-DFT ones. As for the previous approaches, the gap widths are highly dependent on the particular molecular arrangement, although a relation with the geometry is not evident. The G$_0$W$_0$ results follow approximately the same trend as the DFT ones. Thus, the widest and narrowest band gaps correspond to cases 11 and 13, respectively, and both are attributable to the aforementioned electrostatic interactions between the organic and inorganic moieties of the system. 

Amongst data from Table~2, many-body effects on cubic MAPbI$_3$ were taken into account by Leguy \textit{et al.} \cite{Leguy-16}, Brivio \textit{et al.} \cite{Brivio-14}, Ahmed \textit{et al.} \cite{Ahmed-15}, Quarti \textit{et al.} \cite{quarti-2016} and Bokdam \textit{et al.} \cite{Bokdam-16}. The former three works report calculations on a single cubic MAPbI$_3$ cell for fixed orientations of the MA molecule: along $\langle 100 \rangle$ \cite{Brivio-14,Ahmed-15} and along $\langle 100 \rangle$, $\langle 110 \rangle$ and $\langle 111 \rangle$ \cite{Leguy-16}. Our results are in a very good agreement with those by Brivio \textit{et al.} \cite{Brivio-14}, who report a SOC-GW bandgap of 1.27~eV for a configuration equivalent to our case 3 (1.29~eV) and are in a fairy agreement with Ahmed \textit{et al.}, who report a gap of 1.48~eV. This small discrepancy could be probably ascribed to the slightly different lattice parameter (6.26 \AA$\;$in Ref. \cite{Ahmed-15}).
On the other hand, Leguy \textit{et al.} \cite{Leguy-16} report band gaps of 1.60~eV, 1.46~eV and 1.52~eV for the MA molecule aligned along $\langle 100 \rangle$, $\langle 110 \rangle$ and $\langle 111 \rangle$, respectively, using the quasi-particle self-consistent scheme for the GW approximation (QSGW) \cite{Schilfgaarde-06}.
Our band gaps for the equivalent orientations are 1.29~eV (case 3), 1.12~eV (case 2) and 1.20~eV (case 1), and they follow the same trend $\varepsilon_g(\langle 100 \rangle$) $>$ $\varepsilon_g(\langle 111 \rangle)$ $>$ $\varepsilon_g(\langle 110 \rangle)$. The difference of 0.3~eV could be ascribed to different solution schemes of the quasi-particle equation (self-consistent QSGW or scGW vs $G_0W_0$), as also observed in Refs.~\cite{Brivio-14,filip-14b} and reported in Table~2. A value of 1.67~eV was obtained by Bokdam \textit{et al.}, who considered self-consistent-eigenvalue GW$_0$ for a conventional cell with the MA molecules aligned along $[001]$~\cite{Bokdam-16}.

Very recently, Quarti \textit{et al.} evaluated the band gap of MAPbI$_3$ 
at SOC-GW for the orthorhombic, tetragonal and cubic phases~\cite{quarti-2016}.
The band gap of the cubic phase was evaluated  considering structures 
obtained by relaxations of all atomic positions or after relaxation of 
those of MA cations only. For these two configurations, they obtained 
band gaps of 1.28~eV and 1.16~eV, respectively. In the same work, the 
authors considered snapshots from Car-Parrinello Molecular Dynamics 
trajectories of 32 MAPbI$_3$ units. They observed, at DFT and SOC-DFT 
levels, a huge variation of the band gap of the system in the scale of few 
picoseconds, which was related to the motion of the organic cations as well 
as to fluctuations of the inorganic cage.
These results point out that the experimental band gap of around 1.67~eV may be reproduced 
only taking into account the time evolution of the system and 
considering an average of the instantaneous gaps resulting from the 
complex evolution of the MA molecules and the inorganic cage. These results 
are consistent with our data from Table~\ref{T4}, that 
show differences over 1~eV, depending on the orientations of the MA 
molecule at all the considered theoretical levels.

\section{Conclusions and outlook}
\label{conclusion}
The efficiency of cubic CH$_3$NH$_3$PbI$_3$ perovskite for solar cells applications depends on its optical and transport properties. In this work, we 
carried out an \textit{ab initio} study of some realistic orientations of the methylammonium molecules within a PbI$_3$ frame made upon eight elementary 
cells. Several levels of approximation, namely DFT, SOC-DFT, pSIC and SOC-GW, were used. 
We found, at all the levels of theory,  that the particular arrangement of the molecules within the inorganic cage induced a remarkable effect on the value of 
the band gap. In particular, a difference of more than 1~eV was observed depending on the particular configuration. The fact that this ``breathing'' effect appears at several levels of approximation points out that the band gap variation is related to distortions of the PbI$_3$ cage which, in turn, are caused by interactions between the C or N ends of the MA molecule and 
the Pb or I atoms of the cage. Additionally, we reveal that the entity of the self-interaction and self-energy corrections are also dependent on the particular molecular configuration.

These findings point to a remarkable effect on the optical properties as well.
Regarding this, the role played by the excitons in MAPbI$_3$ is another 
controversial issue. Frost \textit{et al.} estimated the binding energy of 
excitons in this system to be about 0.7 meV \--too small to affect 
significantly the photovoltaic performance~\cite{Frost-14}. This is in contradiction with experimental 
evidences that indicate a value in the range 6-55~meV\cite{Ishihara-94,Green-15,tanaka-2003,miyata-2015,Innocenzo-14,savenije-2014,Kitazawa-02,Leguy-16}. 
Solutions of the Bethe-Salpeter equation 
for a single molecular orientation revealed huge exciton binding energies of 153~meV for the cubic phase~\cite{Ahmed-15} and 40~meV for the tetragonal phase~\cite{Zhu-14}.
These high binding energies have been recently rationalized by 
Bokdam \textit{et al.} \cite{Bokdam-16}, who showed that a very dense $k-$point sampling is required to converge the optical spectra for this system; the authors obtained a converged value of 45~meV for the pseudo cubic phase, in better agreement with the experiments.

Our results indicate a strong influence of the orientation of the MA molecules on the electronic structure of MAPbI$_3$, and the question arises as to whether the same influence exists on optical properties. Such an effect was recently studied by Bokdam \textit{et al.} \cite{Bokdam-16}, who considered
a $\sqrt 2\times\sqrt 2$ supercell containing 2 MA molecules 
with different orientations;
a raise of 6~meV was observed when compared with the binding energy calculated with conventional cell. Experimental studies have demonstrated that the onset of the optical absorption spectrum of tetragonal and orthorhombic MAPbI$_3$ shifts to higher energies as the temperatures increases. Also the Urbach energy increases, causing the optical spectrum to become less steep, with temperature\cite{singh-16,Innocenzo-14,sestu-15}. Both facts suggest that the optical gap and the Urbach energy increase with the degree of disorder of the system. Although no data for the cubic phase are available, it is likely that the same effect remains valid. Our results are then consistent with the experimental evidence. Indeed, as the temperature increases, the molecules in the system arrange themselves in orientations with higher formation energies. This would yield blue-shifted optical absorption spectra onsets, as shown in Table~\ref{T1}, and likely higher Urbach energies as well. We cannot discuss in further detail the effect of the MA orientation on the optical properties, as this would require to use more sophisticated theoretical methods (in particular, the calculation of the optical spectrum for each molecular orientation) which are beyond the aim of this work. In addition, the effect of temperature cannot be taken into account in our \textit{ab initio} formalism. In any case, the results reported herein suggest an effect of the molecular arrangement on the optical properties of the system.

\section*{Acknowledgement}
Calculations have been performed in the Cyfronet Computer Centre using the Prometheus computer, 
which is a part of the PL-Grid Infrastructure, and the Kepler cluster, 
from the Department of Physics of the University of Extremadura in Spain. 
This work has been supported by The National Science Centre of Poland 
(the Project No. 2013/11/B/ST3/04041), by the Junta of Extremadura through Grant GR15105. 
DV acknowledges partial support from the EU Centre of Excellence 
"MaX - Materials Design at the Exascale" (Grant No. 676598). 
The authors are greatly thankful to Mr. Maciej Czuchry, from the Cyfronet Computer Center, 
for his inestimable computational support.

\section*{References}

\bibliography{bib}

\end{document}